\documentclass[prd,aps,twocolumn,superscriptaddress,nofootinbib,showpacs,10pt,preprintnumbers]{revtex4-1}

\usepackage{amsmath,amssymb,bm,multirow,float,bbold,hyperref}
\usepackage{graphicx}
\usepackage[american]{babel}
\usepackage{enumerate}
\usepackage{xcolor,colortbl}
\usepackage{mathtools}
\usepackage{booktabs}
\usepackage{subcaption}
\usepackage{soul}
\setstcolor{red}
\usepackage{makebox}
\usepackage{multirow}
\definecolor{midblue}{rgb}{0.0, 0.5, 0.0}

\allowdisplaybreaks

\newcommand\varpm{\mathbin{\vcenter{\hbox{%
  \oalign{\hfil$\scriptstyle\hspace{-0.2ex}+\hspace{-0.2ex}$\hfil\cr
          \noalign{\kern-.5ex}
          $\scriptscriptstyle({-})$\cr}%
}}}}

\newcommand\varmp{\mathbin{\vcenter{\hbox{%
  \oalign{\hfil$\scriptstyle\hspace{-0.2ex}-\hspace{-0.2ex}$\hfil\cr
          \noalign{\kern-.5ex}
          $\scriptscriptstyle({+})$\cr}%
}}}}

\begin{document}

\title{Anomaly-free 2HDMs with a gauged abelian symmetry and two generations of right-handed neutrinos} 

\author{Astrid Ordell}\thanks{astrid.ordell@thep.lu.se}
 \affiliation{Department of Astronomy and Theoretical Physics, Lund University, SE-223 62 Lund, Sweden}
 
\author{Roman Pasechnik}\thanks{roman.pasechnik@thep.lu.se}
 \affiliation{Department of Astronomy and Theoretical Physics, Lund University, SE-223 62 Lund, Sweden}

\author{Hugo Ser\^odio}\thanks{hugo.serodio@thep.lu.se}
\affiliation{Department of Astronomy and Theoretical Physics, Lund University, SE-223 62 Lund, Sweden}

\date{\today}

\preprint{LU-TP 20-28}

\begin{abstract}{
Despite the popularity of two-Higgs-doublet models (2HDMs) with a gauged abelian flavor symmetry, the allowed Yukawa textures have only been partly explored so far. In this work, we classify and compare every anomaly-free instance, in the case of having two generations of
right-handed neutrinos and a type-I seesaw mechanism. We found 16 valid implementations in total,
out of which 11 agree well with current experimental bounds. To our knowledge, neither of these models have been considered previously.
}
\end{abstract}



\maketitle

\section{Introduction}\label{s:intro}

Despite there being relatively strict bounds on Higgs couplings to the SM gauge bosons and heavy fermions \cite{Tanabashi:2018oca}, the Higgs sector remains among the least constrained in the SM. The simplest example of a non-minimal Higgs sector is built by adding one extra Higgs doublet and is known 
as the 2HDM (for a detailed review, see e.g.~Refs.~\cite{Branco:1999fs,Branco:2011iw,Ivanov:2017dad}). 

The 2HDM was first proposed by T.~D.~Lee, over 40 years ago, to motivate spontaneous CP violation \cite{PhysRevD.8.1226} and has evolved into one of the most 
well-explored extensions of the SM today. A generic 2HDM (i.e.~one without any imposed symmetries) is, however, not very enlightening, 
as its scalar potential and Yukawa sector contains a large number of free parameters.
Besides, the generic 2HDM features flavor-changing neutral currents (FCNCs) at tree-level, which are heavily constrained by experiments \cite{Crivellin:2013wna}. 

By imposing a continuous or discrete abelian symmetry, one
can convert a 2HDM into a more predictive framework, with the possible realizations for the quark sector presented in Ref.~\cite{Ferreira:2010ir} and confirmed in Ref.~\cite{Ivanov:2013bka,Serodio:2013gka}. One particularly well-explored path is to look for models that avoid tree-level FCNCs by construction. The simplest example of this is 
when all right-handed fermions of the same charge couple to only one of the Higgs doublets, by the implementation of a discrete $\mathbb{Z}_2$ symmetry, and are referred to as being naturally flavor conserving (NFC) \cite{Glashow1977:M1,Paschos1977}. 
Another, softer, method of avoiding the tree-level FCNCs, is by imposing the Yukawa couplings of the two scalar doublets to align in flavor space, as done in the so-called flavor-aligned 2HDM \cite{PhysRevD.80.091702,Jung:2010aa,Ferreira:2010xe}.

Rather than entirely forbidding tree-level FCNCs, an alternative procedure is to suppress 
them sufficiently to comply with experimental data. A particularly simple and important scenario of such a minimally flavor 
violating (MFV) scheme is realized in Branco, Grimus and Lavoura's (BGL) model \cite{Branco:1996bq,Botella:2014ska}.
In Ref.~\cite{Botella:2011ne} such a BGL formulation was 
extended to incorporate a mechanism for generating neutrino masses, while a gauged version of the BGL model was
implemented for the first time in Ref.~\cite{Celis:2015ara}.

Due to their simplicity and elegance, most of the available models on the market today realize either NFC or MFV scenarios. While this is a reasonable outcome, their dominance has led to a significant research gap in terms of possible Yukawa textures. For example, as noted in our earlier work \cite{Astrid:2019}, more than half of all anomaly-free implementations of a 2HDM with a gauged $U(1)$ flavor symmetry had never been previously considered. 

Another challenge in the physics community has been to explain the smallness of neutrino masses \cite{MINKOWSKI1977421,GellMann:1980vs,Glashow:1979nm,Sawada:1979dis,Ma:2006km,Davidson:2009ha,Gabriel:2006ns,Haba:2011aa,PhysRevD.22.2227,PhysRevD.25.774}.
In this paper, we again classify and compare all anomaly-free implementations of a 2HDM with a gauged abelian flavor symmetry, but this time around also incorporating a mechanism for neutrino mass generation. In particular, we consider a type-I seesaw mechanism\cite{MINKOWSKI1977421,GellMann:1980vs,Glashow:1979nm,Sawada:1979dis} with two generations of right-handed neutrinos \cite{Barreiros:2018ndn,Frampton:2002qc,Ibarra:2003up,Harigaya:2012bw,Rink:2016vvl,Shimizu:2017fgu}.\footnote{We have also classified all anomaly-free models with three generations of right-handed neutrinos. However, with it resulting in over 200 valid implementations, it was no longer feasible to present and compare them all, which is why this paper is limited to the two-generation case.} This corresponds to having two massive and one massless generation of left-handed neutrinos, which is the minimal possible setting in agreement with current experimental bounds \cite{Tanabashi:2018oca}. From this, we found that 11 models, out of the total 16, were in good agreement with data, with three of them being particularly promising.  


The paper is organized as follows: In Sec.~\ref{s:Lag} and \ref{s:Procedure}, we introduce our model and the method of finding anomaly-free solutions, respectively. Next, in Sec.~\ref{s:AnomaFree}, we present the explicit form of all anomaly-free implementations, and, finally, Sec.~\ref{s:Scan}, \ref{s:Pheno} and \ref{s:Concl} involves the scan, its results and our conclusions.

\section{Model specifics}\label{s:Lag}

In this paper, we consider models where the SM particle content is extended by two generations of right-handed neutrinos, a Higgs doublet $\Phi$, a $Z^\prime$ gauge boson and a scalar singlet $S$. In general, all matter-sectors (SM and non-SM) are charged under the new $U(1)^\prime$ symmetry and transform as 
\begin{align}
\label{eq:AbelianTransform}
\begin{split}
\chi \rightarrow e^{i\alpha X_{\chi}}{\chi},
\end{split}
\end{align}
for $\chi=\{{q_\mathrm{L}^0}_j,\;{d_\mathrm{R}^0}_j,\;{u_\mathrm{R}^0}_j,\;{\ell_\mathrm{L}^0}_j,\;{e_\mathrm{R}^0}_j,\; {\nu_\mathrm{R}^0}_j, \;\Phi_{a}, \;S\}$, where $X_{\chi}$ is the corresponding $U(1)^\prime$ charge, which is in general flavor dependent, where  $j$ is a flavor index, $a=1,2$, and where the zero superscript denotes the flavor basis. 

While a subset of the charges can be set to zero, we always require that the singlet, and at least one of the Higgs doublets, are charged under $U(1)^\prime$. As such, the scalar masses are not solely determined by the EW scale and by the dimensionless couplings in the scalar potential (bounded by unitarity), but can be adjusted via the VEV of the scalar singlet. 

In the flavor basis, the Yukawa interactions are then given by

\begin{align}
\begin{split}
-\mathcal{L}_\mathrm{Yukawa}&=\overline{q_L^0}\Gamma_a\Phi_a d_R^0+\overline{q_L^0}\Delta_a\tilde{\Phi}_a u_R^0\\
&+\overline{\ell_L^0}\Pi_a\Phi_a e_R^0+\overline{\ell_L^0}\Sigma_a\tilde{\Phi}_a \nu_R\\
&+\frac{1}{2}\overline{\nu_R^c}\left({{A}}+{{B}}S+{{C}}S^\ast\right)\nu_R+\mathrm{H.c.}\,,
\end{split}
\end{align}   
{where $\Gamma_a$, $\Delta_a$, $\Pi_a$ and $\Sigma_a$ are the Yukawa couplings for down-type quarks, up-type quarks, charged leptons and Dirac neutrinos, respectively, and where the matrix $A$ is the bare neutrino mass, while $B$ and $C$ are Majorana-like Yukawa couplings. The scalar fields are parameterized as} 

\begin{align}
\begin{split}
\label{eq:Scalars}
\Phi_a &= \frac{1}{\sqrt{2}}
\begin{pmatrix}
\sqrt{2}\phi^+_a\\
v_a\,e^{i\alpha_a}+R_a+iI_a
\end{pmatrix}\,,\\[0.8mm]
S &=\frac{1}{\sqrt{2}} \left(v_S\,e^{i\alpha_S}+\rho+i\eta\right)\,,
\end{split}
\end{align}
where $v_S$ and $v_a$, $a=1,2$, are the VEVs of the scalar singlet and the Higgs doublets, respectively. With the additional abelian symmetry, we can set two 
of the VEVs real, e.g.~$\alpha_1=\alpha_2=0$. 

After EW symmetry breaking, the mass matrices for the quarks and charged leptons are given by 

\begin{align}
\begin{split}
M_u&=\frac{1}{\sqrt{2}}\left(v_1 \Delta_1+v_2\Delta_2\right), 
\end{split}
\end{align}
and equivalently for $M_d$ and $M_e$, but with $\Delta$ exchanged for $\Gamma$ and $\Pi$, respectively. For the neutrino sector, on the other hand, we have

\begin{align}
\begin{split}
-2\mathcal{L}_{\nu}^\mathrm{mass}=&
\begin{pmatrix}
\overline{\nu_L^0}&\overline{\nu_R^c}
\end{pmatrix}
\begin{pmatrix}
\mathbb{0}&m_D\\
m_D^T&M_R
\end{pmatrix}
\begin{pmatrix}
\nu_L^{0,c}\\
\nu_R
\end{pmatrix}+\text{H.c.}\,,
\end{split}
\end{align}

with 

\begin{align}
\begin{split}
m_D&=\frac{1}{\sqrt{2}}\left(v_1\Sigma_1+v_2\Sigma_2\right),\\
M_R&={{A}}+\frac{v_S}{\sqrt{2}}\left({{B}}\,e^{i\alpha_S}+{{C}}\,e^{-i\alpha_S}\right).
\end{split}
\end{align}

In the limit of $M_R\gg m_D$, the light eigenstates are the left-handed Majorana particles, such that the effective mass term is given by 

\begin{align}
\label{eq:MnuDef}
\begin{split}
2\mathcal{L}_\nu^\mathrm{eff}=\overline{\nu_L^0}\left(m_D M_R^{-1} m_D^T\right) \nu_L^{0,c}+\text{H.c.}\;.
\end{split}
\end{align}

As the heavy eigenstates decouple, the scalar potential and the new gauge sector have the same form as in  Ref.~\cite{Astrid:2019}.

\section{Procedure for finding anomaly-free solutions}\label{s:Procedure}

For two generations of massive neutrinos, we have the following physical requirements in the leptonic sector

\begin{itemize}
\item[(i)] No massless charged leptons, i.e.~$\text{det}\hspace{0.5mm}M_e\neq 0$;
\item[(ii)] Two generations of massive neutrinos, i.e. $\text{rank} \hspace{0.5mm} M_\nu= 2$;
\item[(iii)] A non-zero complex phase in the PMNS matrix, i.e. $\text{det}\hspace{0.5mm}[M_eM_e^\dagger,M_\nu M_\nu^\dagger]\neq 0$;
\end{itemize}
with $M_\nu$ defined as $-m_D M_R^{-1} m_D^T$, as apparent from Eq.~\eqref{eq:MnuDef}. With the neutrinos gaining their mass via a type-I seesaw mechanism, condition number (ii) corresponds to $M_\mathrm{R}$ being a $2\times2$ matrix with a non-zero determinant and $m_\mathrm{D}$ being a $3\times2$ matrix with rank 2.

The key is to find an efficient method for identifying all textures that fulfill these constraints, as naively considering every possible combination would require an immense amount of computational power. To this end, we will use the procedure introduced in Ref.~\cite{Astrid:2019}, i.e.~using a set of \emph{minimal constraints}, in addition to avoiding textures that are equivalent up to permutations of flavor indices. 

\subsection{The charged lepton sector}

Let us begin with the charged lepton sector, which is the only sector with the same minimal textures as in Ref.~\cite{Astrid:2019}. Here, the requirement of a non-zero determinant forces the combined texture of $\Pi_1$ and $\Pi_2$ to have at least one non-zero entry in each row and each column. In other words, there are six possible textures for $M_e$ -- all being permutations of the diagonal texture

\begin{equation}
\label{Me}
M_e:
\begin{pmatrix}
 \times &   \makebox[\widthof{$\times$}][c]{0}&  \makebox[\widthof{$\times$}][c]{0}
\\
  \makebox[\widthof{$\times$}][c]{0} &  \times&   \makebox[\widthof{$\times$}][c]{0}
\\
 \makebox[\widthof{$\times$}][c]{0} &   \makebox[\widthof{$\times$}][c]{0}&   \times
\end{pmatrix}.
\end{equation}

As we have not yet introduced the sectors involving Dirac neutrinos, Majorana neutrinos and quarks, permutations of row and columns simply amount to a relabelling of flavor indices. Hence, we only need to consider one of the possible six textures,  e.g.~the diagonal one presented in Eq.~\eqref{Me}. This leaves four possibilities for the minimal textures of $\Pi_1$ and $\Pi_2$, namely

\begin{align}
\label{texturesLepton}
\begin{split}
(1)\;\;\;\;\;\;\Pi_1:
\begin{pmatrix}
\times &   \makebox[\widthof{$\times$}][c]{0}&  \makebox[\widthof{$\times$}][c]{0}
\\
 \makebox[\widthof{$\times$}][c]{0} &   \times&   \makebox[\widthof{$\times$}][c]{0}
\\
 \makebox[\widthof{$\times$}][c]{0} &   \makebox[\widthof{$\times$}][c]{0}&    \times
\end{pmatrix},
\;\;
\Pi_2:
\begin{pmatrix}
 \makebox[\widthof{$\times$}][c]{0}&   \makebox[\widthof{$\times$}][c]{0}&  \makebox[\widthof{$\times$}][c]{0}
\\
  \makebox[\widthof{$\times$}][c]{0} &  \makebox[\widthof{$\times$}][c]{0}&   \makebox[\widthof{$\times$}][c]{0}
\\
 \makebox[\widthof{$\times$}][c]{0} &   \makebox[\widthof{$\times$}][c]{0}&   \makebox[\widthof{$\times$}][c]{0}
\end{pmatrix},
\\
(2)\;\;\;\;\;\;\Pi_1:
\begin{pmatrix}
 \times &   \makebox[\widthof{$\times$}][c]{0}&  \makebox[\widthof{$\times$}][c]{0}
\\
  \makebox[\widthof{$\times$}][c]{0} &   \times&   \makebox[\widthof{$\times$}][c]{0}
\\
 \makebox[\widthof{$\times$}][c]{0} &   \makebox[\widthof{$\times$}][c]{0}&    \makebox[\widthof{$\times$}][c]{0}
\end{pmatrix},
\;\;
\Pi_2:
\begin{pmatrix}
 \makebox[\widthof{$\times$}][c]{0}&   \makebox[\widthof{$\times$}][c]{0}&  \makebox[\widthof{$\times$}][c]{0}
\\
  \makebox[\widthof{$\times$}][c]{0} & \makebox[\widthof{$\times$}][c]{0}&   \makebox[\widthof{$\times$}][c]{0}
\\
 \makebox[\widthof{$\times$}][c]{0} &   \makebox[\widthof{$\times$}][c]{0}&   \times
\end{pmatrix},
\\
(3)\;\;\;\;\;\;\Pi_1:
\begin{pmatrix}
 \times &   \makebox[\widthof{$\times$}][c]{0}&  \makebox[\widthof{$\times$}][c]{0}
\\
  \makebox[\widthof{$\times$}][c]{0} & \makebox[\widthof{$\times$}][c]{0}&   \makebox[\widthof{$\times$}][c]{0}
\\
 \makebox[\widthof{$\times$}][c]{0} &   \makebox[\widthof{$\times$}][c]{0}&    \makebox[\widthof{$\times$}][c]{0}
\end{pmatrix},
\;\;
\Pi_2:
\begin{pmatrix}
 \makebox[\widthof{$\times$}][c]{0}&   \makebox[\widthof{$\times$}][c]{0}&  \makebox[\widthof{$\times$}][c]{0}
\\
  \makebox[\widthof{$\times$}][c]{0} &  \times&   \makebox[\widthof{$\times$}][c]{0}
\\
 \makebox[\widthof{$\times$}][c]{0} &   \makebox[\widthof{$\times$}][c]{0}&   \times
\end{pmatrix},
\\
(4)\;\;\;\;\;\;\Pi_1:
\begin{pmatrix}
 \makebox[\widthof{$\times$}][c]{0} &   \makebox[\widthof{$\times$}][c]{0}&  \makebox[\widthof{$\times$}][c]{0}
\\
  \makebox[\widthof{$\times$}][c]{0} & \makebox[\widthof{$\times$}][c]{0}&   \makebox[\widthof{$\times$}][c]{0}
\\
 \makebox[\widthof{$\times$}][c]{0} &   \makebox[\widthof{$\times$}][c]{0}&    \makebox[\widthof{$\times$}][c]{0}
\end{pmatrix},
\;\;
\Pi_2:
\begin{pmatrix}
  \times &   \makebox[\widthof{$\times$}][c]{0}&  \makebox[\widthof{$\times$}][c]{0}
\\
  \makebox[\widthof{$\times$}][c]{0} &  \times&   \makebox[\widthof{$\times$}][c]{0}
\\
 \makebox[\widthof{$\times$}][c]{0} &   \makebox[\widthof{$\times$}][c]{0}&   \times
\end{pmatrix}.
\end{split}
\end{align}
Note that, as these are so-called minimal textures, additional non-zero entries are allowed.  

The textures are then translated into a set of linear equations for the $U(1)^\prime$ charges, since, 
in order for the abelian flavor symmetry in Eq.~\eqref{eq:AbelianTransform} to be a symmetry of the Lagrangian, it is required that

\begin{equation}
\left(\Pi_a\right)_{ij}=e^{i\alpha(X_{\ell_i}-X_{e_j}-X_{\Phi_a})}\left(\Pi_a\right)_{ij},
\end{equation} 
which corresponds to

\begin{align}
\begin{split}
\left(\Pi_a\right)_{ij}&=\mathrm{any\;\;\;\; if\;\;} X_{\ell_i}-X_{e_j}=X_{\Phi_a},
\\
\left(\Pi_a\right)_{ij}&=0\mathrm{\;\;\;\;\;\;\;\; if\;\;} X_{\ell_i}-X_{e_j}\neq X_{\Phi_a},
\end{split}
\end{align} 
where $i,j$ are the flavor indices running from one to three, and where there is no summation over the repeated indices. When converting the textures in Eq.~\eqref{texturesLepton}, we exclude constraints on the form $X_{\ell_i}-X_{e_j}\neq X_{\Phi_a}$, {as we want to} allow for additional non-zero entries. As such, the textures correspond to

\begin{align}
\begin{split}
\label{LeptonConstraints}
X_{\ell_{1(2)}}-X_{e_{1(2)}}=X_{\Phi_1},&\hspace{6.8mm}X_{\ell_i}-X_{e_i}=X_{\Phi_1},
\\[0.7mm]
X_{\ell_{3}}-X_{e_{3}}=X_{\Phi_2},&\hspace{6.5mm}X_{\ell_1}-X_{e_1}=X_{\Phi_1},
\\[0.7mm]
X_{\ell_{2(3)}}-X_{e_{2(3)}}=X_{\Phi_2},&\hspace{7.3mm}X_{\ell_i}-X_{e_i}=X_{\Phi_2}.
\end{split}
\end{align}

Note that, in excluding constraints on the form $X_{\ell_i}-X_{e_j}\neq X_{\Phi_a}$, the conditions in Eq.~\eqref{LeptonConstraints} are sufficient for generating \emph{all} viable lepton textures.

\subsection{The Majorana neutrinos}
 
For the Majorana neutrinos, there are nine possible minimal textures for $A$, $B$ and $C$ that fulfill the constraints of $M_\mathrm{R}$ being symmetric and having a non-zero determinant. Once again, we do not need to consider any permutations of rows and columns, as these permutations are separate from the ones in the charged lepton sector, and hence simply amount to a relabelling of indices. The first three minimal textures are given by 

\begin{align}
\begin{split}
\label{A}
(1)&\;\;\;\;A:
\begin{pmatrix}
\makebox[\widthof{$\times$}][c]{0}&  \times
\\
\times & \makebox[\widthof{$\times$}][c]{0}
\end{pmatrix},
\;\;
B:
\begin{pmatrix}
\makebox[\widthof{$\times$}][c]{0}& \makebox[\widthof{$\times$}][c]{0}
\\
\makebox[\widthof{$\times$}][c]{0}& \makebox[\widthof{$\times$}][c]{0}
\end{pmatrix},
\;\;
C:
\begin{pmatrix}
\makebox[\widthof{$\times$}][c]{0}&\makebox[\widthof{$\times$}][c]{0}
\\
\makebox[\widthof{$\times$}][c]{0} & \makebox[\widthof{$\times$}][c]{0}
\end{pmatrix},
\\
(2)&\;\;\;\;A:
\begin{pmatrix}
\makebox[\widthof{$\times$}][c]{0}& \makebox[\widthof{$\times$}][c]{0}
\\[1mm]
\makebox[\widthof{$\times$}][c]{0}& \makebox[\widthof{$\times$}][c]{0}
\end{pmatrix},
\;\;
B:
\begin{pmatrix}
\makebox[\widthof{$\times$}][c]{0}&  \times
\\
\times & \makebox[\widthof{$\times$}][c]{0}
\end{pmatrix},
\;\;
C:
\begin{pmatrix}
\makebox[\widthof{$\times$}][c]{0}&\makebox[\widthof{$\times$}][c]{0}
\\
\makebox[\widthof{$\times$}][c]{0} & \makebox[\widthof{$\times$}][c]{0}
\end{pmatrix},
\\[1mm]
(3)&\;\;\;\;A:
\begin{pmatrix}
\makebox[\widthof{$\times$}][c]{0}& \makebox[\widthof{$\times$}][c]{0}
\\
\makebox[\widthof{$\times$}][c]{0}& \makebox[\widthof{$\times$}][c]{0}
\end{pmatrix},
\;\;
B:
\begin{pmatrix}
\makebox[\widthof{$\times$}][c]{0}& \makebox[\widthof{$\times$}][c]{0}
\\
\makebox[\widthof{$\times$}][c]{0}& \makebox[\widthof{$\times$}][c]{0}
\end{pmatrix},
\;\;
C:
\begin{pmatrix}
\makebox[\widthof{$\times$}][c]{0}&  \times
\\
\times & \makebox[\widthof{$\times$}][c]{0}
\end{pmatrix},
\end{split}
\end{align}
which corresponds to the constraints $X_{\nu_1}+X_{\nu_2}=0$, $X_{\nu_1}+X_{\nu_2}=-X_S$ and $X_{\nu_1}+X_{\nu_2}=X_S$, respectively, since invariance under the flavor symmetry requires that 

\begin{align}
\begin{split}
A_{ij}&=e^{i\alpha(X_{\nu_i}+X_{\nu_j})}A_{ij},
\\
B_{ij}&=e^{i\alpha(X_{\nu_i}+X_{\nu_j}+X_S)}B_{ij},
\\
C_{ij}&=e^{i\alpha(X_{\nu_i}+X_{\nu_j}-X_S)}C_{ij}.
\end{split}
\end{align}
Again, there are no constraints in the form of $2X_{\nu_{1(2)}}\neq 0$ or  $2X_{\nu_{1(2)}}\neq \pm X_S$ and, as such, the final textures are not limited to the ones in Eq.~\ref{A}. 

Next, we have the remaining six minimal textures for the Majorana neutrinos, namely

\begin{align}
\begin{split}
\label{B}
1:
\begin{pmatrix}
\times&\makebox[\widthof{$\times$}][c]{0}
\\
\makebox[\widthof{$\times$}][c]{0} & \makebox[\widthof{$\times$}][c]{0}
\end{pmatrix},
\;\;
2:
\begin{pmatrix}
\makebox[\widthof{$\times$}][c]{0}&\makebox[\widthof{$\times$}][c]{0}
\\
\makebox[\widthof{$\times$}][c]{0} & \times
\end{pmatrix},
\;\;
3:
\begin{pmatrix}
\makebox[\widthof{$\times$}][c]{0}&  \makebox[\widthof{$\times$}][c]{0}
\\
\makebox[\widthof{$\times$}][c]{0} & \makebox[\widthof{$\times$}][c]{0}
\end{pmatrix},
\end{split}
\end{align}
where $\{1,2,3\}$ can be any permutation of $\{A,B,C\}$. 
The corresponding constraints are given by 

\begin{align}
\begin{split}
2X_{\nu_2}=-X_S,&\hspace{8mm}2X_{\nu_1}=0 \;\;\;\mathrm{for}\;\;\;\{A,B,C\},
\\
2X_{\nu_1}=-X_S,& \hspace{8mm}2X_{\nu_2}=0\;\;\;\mathrm{for}\;\;\;\{B,A,C\},
\\
2X_{\nu_2}=X_S,& \hspace{8mm}2X_{\nu_1}=0\;\;\;\mathrm{for}\;\;\;\{A,C,B\},
\\
2X_{\nu_1}=X_S,&\hspace{8mm}2X_{\nu_2}=0\;\;\;\mathrm{for}\;\;\;\{C,A,B\},
\\
2X_{\nu_1}=-X_S,& \hspace{5mm}2X_{\nu_2}=X_S\;\;\;\mathrm{for}\;\;\;\{B,C,A\},
\\
2X_{\nu_2}=-X_S,& \hspace{5mm}2X_{\nu_1}=X_S\;\;\;\mathrm{for}\;\;\;\{C,B,A\}.
\end{split}
\end{align}

Furthermore, invariance under the phase sensitive part of the scalar potential in Appendix \ref{append:scalar}, requires for one of the following four constraints to be met


%

\begin{align}
\label{ScalarPot}
\begin{split}
X_S=\pm\left(X_{\Phi_1}-X_{\Phi_2}\right), \;\;X_S=\pm\frac{1}{2}\left(X_{\Phi_1}-X_{\Phi_2}\right).
\end{split}
\end{align}

Note that, in cases where only $A$ has a non-zero texture, all four solutions are allowed.

\subsection{The Dirac neutrinos}

Next, we have the possible textures for Dirac neutrinos, where the permutations of rows are not independent of those in the charged lepton sector, and where permutation of columns are not separate from those in the Majorana neutrino sector. As such, we must consider every possible permutation explicitly, which for a rank 2 $m_\mathrm{D}$ corresponds to a total of 6 possible textures

\begin{align}
\begin{split}
\label{Dirac}
1:
\begin{pmatrix}
\times&\makebox[\widthof{$\times$}][c]{0}
\\
\makebox[\widthof{$\times$}][c]{0} & \times
\\
\makebox[\widthof{$\times$}][c]{0} & \makebox[\widthof{$\times$}][c]{0}
\end{pmatrix},
\;\;
2:
\begin{pmatrix}
\times&\makebox[\widthof{$\times$}][c]{0}
\\
\makebox[\widthof{$\times$}][c]{0} & \makebox[\widthof{$\times$}][c]{0}
\\
\makebox[\widthof{$\times$}][c]{0} & \times
\end{pmatrix},
\;\;
3:
\begin{pmatrix}
\makebox[\widthof{$\times$}][c]{0}&\makebox[\widthof{$\times$}][c]{0}
\\
 \times & \makebox[\widthof{$\times$}][c]{0}
\\
\makebox[\widthof{$\times$}][c]{0} & \times
\end{pmatrix},
\\
4:
\begin{pmatrix}
\makebox[\widthof{$\times$}][c]{0}&\times
\\
\times &\makebox[\widthof{$\times$}][c]{0}
\\
\makebox[\widthof{$\times$}][c]{0} & \makebox[\widthof{$\times$}][c]{0}
\end{pmatrix},
\;\;
5:
\begin{pmatrix}
\makebox[\widthof{$\times$}][c]{0}& \times
\\
\makebox[\widthof{$\times$}][c]{0} & \makebox[\widthof{$\times$}][c]{0}
\\
 \times &\makebox[\widthof{$\times$}][c]{0}
\end{pmatrix},
\;\;
6:
\begin{pmatrix}
\makebox[\widthof{$\times$}][c]{0}&\makebox[\widthof{$\times$}][c]{0}
\\
\makebox[\widthof{$\times$}][c]{0} & \times
\\
 \times &\makebox[\widthof{$\times$}][c]{0}
\end{pmatrix},
\end{split}
\end{align}
where each matrix comes in four versions, depending on whether the non-zero texture sits in $\Sigma_1$ or $\Sigma_2$. For example, for texture number 1 we have the four possibilities

\begin{align}
\begin{split}
\label{DiracExample}
(\mathrm{i})\;\;\;\;\;\;\Sigma_1:
\begin{pmatrix}
 \times &\makebox[\widthof{$\times$}][c]{0}
\\
\makebox[\widthof{$\times$}][c]{0} & \times
\\
\makebox[\widthof{$\times$}][c]{0} & \makebox[\widthof{$\times$}][c]{0}
\end{pmatrix},
\;\;
\Sigma_2:
\begin{pmatrix}
\makebox[\widthof{$\times$}][c]{0}&\makebox[\widthof{$\times$}][c]{0}
\\
\makebox[\widthof{$\times$}][c]{0} &\makebox[\widthof{$\times$}][c]{0}
\\
\makebox[\widthof{$\times$}][c]{0} & \makebox[\widthof{$\times$}][c]{0}
\end{pmatrix},
\\
(\mathrm{ii})\;\;\;\;\;\;\Sigma_1:
\begin{pmatrix}
 \times &\makebox[\widthof{$\times$}][c]{0}
\\
\makebox[\widthof{$\times$}][c]{0} & \makebox[\widthof{$\times$}][c]{0}
\\
\makebox[\widthof{$\times$}][c]{0} & \makebox[\widthof{$\times$}][c]{0}
\end{pmatrix},
\;\;
\Sigma_2:
\begin{pmatrix}
\makebox[\widthof{$\times$}][c]{0}&\makebox[\widthof{$\times$}][c]{0}
\\
\makebox[\widthof{$\times$}][c]{0} & \times 
\\
\makebox[\widthof{$\times$}][c]{0} & \makebox[\widthof{$\times$}][c]{0}
\end{pmatrix},
\\
(\mathrm{iii})\;\;\;\;\;\;\Sigma_1:
\begin{pmatrix}
\makebox[\widthof{$\times$}][c]{0}&\makebox[\widthof{$\times$}][c]{0}
\\
\makebox[\widthof{$\times$}][c]{0} & \times
\\
\makebox[\widthof{$\times$}][c]{0} & \makebox[\widthof{$\times$}][c]{0}
\end{pmatrix},
\;\;
\Sigma_2:
\begin{pmatrix}
\times&\makebox[\widthof{$\times$}][c]{0}
\\
\makebox[\widthof{$\times$}][c]{0} &\makebox[\widthof{$\times$}][c]{0}
\\
\makebox[\widthof{$\times$}][c]{0} & \makebox[\widthof{$\times$}][c]{0}
\end{pmatrix},
\\
(\mathrm{iv})\;\;\;\;\;\;\Sigma_1:
\begin{pmatrix}
\makebox[\widthof{$\times$}][c]{0}&\makebox[\widthof{$\times$}][c]{0}
\\
\makebox[\widthof{$\times$}][c]{0} &\makebox[\widthof{$\times$}][c]{0}
\\
\makebox[\widthof{$\times$}][c]{0} & \makebox[\widthof{$\times$}][c]{0}
\end{pmatrix},
\;\;
\Sigma_2:
\begin{pmatrix}
 \times &\makebox[\widthof{$\times$}][c]{0}
\\
\makebox[\widthof{$\times$}][c]{0} & \times
\\
\makebox[\widthof{$\times$}][c]{0} & \makebox[\widthof{$\times$}][c]{0}
\end{pmatrix},
\end{split}
\end{align}
with the corresponding constraints given by

\begin{align}
\begin{split}
\label{DiracConstraints}
(\mathrm{i})\;\;-X_{\ell_1}+X_{\nu_1}=X_{\Phi_1}, \; -X_{\ell_2}+X_{\nu_2}=X_{\Phi_1},
\\[0.7mm]
(\mathrm{ii})\;\;-X_{\ell_1}+X_{\nu_1}=X_{\Phi_1}, \; -X_{\ell_2}+X_{\nu_2}=X_{\Phi_2},
\\[0.7mm]
(\mathrm{iii})\;\;-X_{\ell_1}+X_{\nu_1}=X_{\Phi_2}, \; -X_{\ell_2}+X_{\nu_2}=X_{\Phi_1},
\\[0.7mm]
(\mathrm{iv})\;\;-X_{\ell_1}+X_{\nu_1}=X_{\Phi_2}, \; -X_{\ell_2}+X_{\nu_2}=X_{\Phi_2},
\end{split}
\end{align}
as invariance under the flavor symmetry requires that

\begin{equation}
\left(\Sigma_a\right)_{ij}=e^{i\alpha(X_{\ell_i}-X_{\nu_j}+X_{\Phi_a})}\left(\Sigma_a\right)_{ij}.
\end{equation} 

\subsection{The quark sector}

For the quark sector, on the other hand, we can use the textures from Ref.~\cite{Ferreira:2010ir} that corresponds to continuous symmetries, rather than repeating the procedure with minimal constraints. The quark textures in Ref.~\cite{Ferreira:2010ir} are then translated into a set of linear constraints using the same method as above, i.e.~by demanding invariance under the flavor symmetry

\begin{align}
\begin{split}
\left(\Gamma_a\right)_{ij}=e^{i\alpha(X_{q_i}-X_{d_j}-X_{\Phi_a})}\left(\Gamma_a\right)_{ij},
\\
\left(\Delta_a\right)_{ij}=e^{i\alpha(X_{q_i}-X_{u_j}+X_{\Phi_a})}\left(\Delta_a\right)_{ij}.
\end{split}
\end{align}

However, this time around, we \emph{do} make use of the non-equalities, i.e.~replacing the zero textures with $X_{q_i}-X_{d_j}\neq X_{\Phi_a}$ and $X_{q_i}-X_{u_j}\neq -X_{\Phi_a}$, respectively, since the result in Ref.~\cite{Ferreira:2010ir} corresponds to complete textures and not to minimal ones. 

\subsection{Combining all sectors}  

When combining the constraints from all sectors above, in addition to including the anomaly cancellation conditions involving $U(1)^\prime$,\footnote{These have the same form as in Ref.~\cite{Astrid:2019}, but with an additional term $-X_{\nu_j}^3$ for $A_{1'1'1'}$, and $-X_{\nu_j}$ for $A_{gg1'}$, respectively, with $j=1,2$. For efficient methods on extracting anomaly-free solutions, see Refs.~\cite{Batra:2005rh,Allanach:2018vjg,Rathsman:2019wyk,PhysRevLett.123.151601,PhysRevD.101.095032}.}  we end up with one large system of equations for the $U(1)^\prime$ charges. In total, there are linear equations coming from the charged lepton textures, Majorana neutrino textures, Dirac neutrino textures, quark textures and from the invariance of the scalar potential, in addition to a mix of linear and non-linear constraints coming from the anomaly constraints. Any rational solution to this system is then labeled as a valid model. Note that, as we loop over every possible combination of constraints, we are guaranteed to find all valid solutions, in comparison to e.g.~a scanning procedure, in which one might only find a subset of them. 

To avoid degenerate solutions once everything is combined, we make sure that neither of the models can be reached from one another via the transformations

\begin{align}
\label{Permutations}
\begin{split}
\Gamma^\prime_{1,2}&=\mathcal{P}_i^\mathrm{T}\Gamma_{1,2} \mathcal{P}_j, \;\;\Delta^\prime_{1,2}=\mathcal{P}_i^\mathrm{T}\Delta_{1,2} \mathcal{P}_k,
\\
\Pi^\prime_{1,2}&=\mathcal{P}_l^\mathrm{T}\Pi_{1,2} \mathcal{P}_m, \;\;\Sigma^\prime_{1,2}=\mathcal{P}_l^\mathrm{T}\Sigma_{1,2} \widetilde{\mathcal{P}}_n,
\\
G^\prime&=\widetilde{\mathcal{P}}_n^\mathrm{T} G \widetilde{\mathcal{P}}_n,
\end{split}
\end{align}
where $\mathcal{P}$ is the three-dimensional representation of the permutation group $S_3$, $\widetilde{\mathcal{P}}$ is the two-dimensional representation of $S_2$ and where $G=A,B,C$. That is, $i,j,k,l,m$ ranges from one to six, while $n$ ranges from one to two.  

\section{Anomaly-free models}\label{s:AnomaFree}
Up to the implementation of Eq.~\eqref{ScalarPot} there are a total of 16 valid models.  Below they are gathered into five separate categories, Category A to E, based on whether they share the same textures. The corresponding charges for each model are presented in Tab.~\ref{tab:charges} and Tab.~\ref{tab:charges2}.

Note that three of the Majorana- and Dirac neutrino textures below (in E3, E4 and E5) have been studied previously in Ref.~\cite{Barreiros:2018ndn}, but in the context of having only the SM particle content.


\subsection{Category A}

Category A consists of two lepton textures and four quark textures, where all eight combinations correspond to valid models, model A1 to A8. The two lepton textures are given by

\begin{align*}
\begin{split}
(\mathrm{E1}):\;\;\;\;\;\;&\Pi_1:
\begin{pmatrix}
\ast&\ast&\makebox[\widthof{$\times$}][c]{0}
\\
\ast&\ast&\makebox[\widthof{$\times$}][c]{0}
\\
\makebox[\widthof{$\times$}][c]{0} & \makebox[\widthof{$\times$}][c]{0} & \times
\end{pmatrix},
\;\;
\Pi_2:
\begin{pmatrix}
\makebox[\widthof{$\times$}][c]{0}&\makebox[\widthof{$\times$}][c]{0}&\makebox[\widthof{$\times$}][c]{0}
\\
\makebox[\widthof{$\times$}][c]{0}&\makebox[\widthof{$\times$}][c]{0}&\makebox[\widthof{$\times$}][c]{0}
\\
 \times& \times&\makebox[\widthof{$\times$}][c]{0} 
\end{pmatrix},
\\
&\Sigma_1:
\begin{pmatrix}
\ast&  \makebox[\widthof{$\times$}][c]{0}
\\
\ast & \makebox[\widthof{$\times$}][c]{0}
\\
\makebox[\widthof{$\times$}][c]{0} & \times
\end{pmatrix},
\;\;
\Sigma_2:
\begin{pmatrix}
 \makebox[\widthof{$\times$}][c]{0}& \times
\\
 \makebox[\widthof{$\times$}][c]{0}& \times
\\
\makebox[\widthof{$\times$}][c]{0} & \makebox[\widthof{$\times$}][c]{0} 
\end{pmatrix},
\\
&A:
\begin{pmatrix}
\times& \makebox[\widthof{$\times$}][c]{0}
\\
\makebox[\widthof{$\times$}][c]{0} & \makebox[\widthof{$\times$}][c]{0} 
\end{pmatrix},
\;\;
B:
\begin{pmatrix}
\makebox[\widthof{$\times$}][c]{0} & \times
\\
\times& \makebox[\widthof{$\times$}][c]{0}
\end{pmatrix},
\;\;
C:
\begin{pmatrix}
\makebox[\widthof{$\times$}][c]{0} & \makebox[\widthof{$\times$}][c]{0}
\\
\makebox[\widthof{$\times$}][c]{0} & \makebox[\widthof{$\times$}][c]{0}
\end{pmatrix},
\end{split}
\end{align*}

\vspace{-2mm}

\begin{align*}
\begin{split}
(\mathrm{E6}):\;\;\;\;\;\;&\Pi_1:
\begin{pmatrix}
\times&\makebox[\widthof{$\times$}][c]{0}&\makebox[\widthof{$\times$}][c]{0}
\\
\makebox[\widthof{$\times$}][c]{0}&\ast&\makebox[\widthof{$\times$}][c]{0}
\\
\makebox[\widthof{$\times$}][c]{0} & \makebox[\widthof{$\times$}][c]{0} & \times
\end{pmatrix},
\;\;
\Pi_2:
\begin{pmatrix}
\makebox[\widthof{$\times$}][c]{0}&\times&\makebox[\widthof{$\times$}][c]{0}
\\
\makebox[\widthof{$\times$}][c]{0}&\makebox[\widthof{$\times$}][c]{0}&\makebox[\widthof{$\times$}][c]{0}
\\
 \times&\makebox[\widthof{$\times$}][c]{0} &\makebox[\widthof{$\times$}][c]{0} 
\end{pmatrix},
\\
&\Sigma_1:
\begin{pmatrix}
\ast&  \times
\\
\makebox[\widthof{$\times$}][c]{0}& \makebox[\widthof{$\times$}][c]{0}
\\
\makebox[\widthof{$\times$}][c]{0} &\makebox[\widthof{$\times$}][c]{0}
\end{pmatrix},
\;\;
\Sigma_2:
\begin{pmatrix}
 \makebox[\widthof{$\times$}][c]{0}& \makebox[\widthof{$\times$}][c]{0}
\\
\ast& \times
\\
\makebox[\widthof{$\times$}][c]{0} & \makebox[\widthof{$\times$}][c]{0} 
\end{pmatrix},
\\
&A:
\begin{pmatrix}
\ast&\times
\\
\times&\times
\end{pmatrix},
\;\;
B:
\begin{pmatrix}
\makebox[\widthof{$\times$}][c]{0} & \makebox[\widthof{$\times$}][c]{0}
\\
\makebox[\widthof{$\times$}][c]{0} & \makebox[\widthof{$\times$}][c]{0}
\end{pmatrix},
\;\;
C:
\begin{pmatrix}
\makebox[\widthof{$\times$}][c]{0} & \makebox[\widthof{$\times$}][c]{0}
\\
\makebox[\widthof{$\times$}][c]{0} & \makebox[\widthof{$\times$}][c]{0}
\end{pmatrix},
\end{split}
\end{align*}
and the four quark textures by

\begin{align*}
\begin{split}
(\mathrm{Q1}):\hspace{6.7mm}&\Gamma_1:
\begin{pmatrix}
\times&\makebox[\widthof{$\times$}][c]{0}&\makebox[\widthof{$\times$}][c]{0}
\\
\makebox[\widthof{$\times$}][c]{0}&\ast&\makebox[\widthof{$\times$}][c]{0}
\\
\makebox[\widthof{$\times$}][c]{0}&\makebox[\widthof{$\times$}][c]{0}&\ast
\end{pmatrix},
\;\;
\Gamma_2:
\begin{pmatrix}
\makebox[\widthof{$\times$}][c]{0}&\makebox[\widthof{$\times$}][c]{0}&\times
\\
\makebox[\widthof{$\times$}][c]{0}&\makebox[\widthof{$\times$}][c]{0}&\makebox[\widthof{$\times$}][c]{0}
\\
\makebox[\widthof{$\times$}][c]{0}&\times&\makebox[\widthof{$\times$}][c]{0} 
\end{pmatrix},
\hspace{6mm}
\\
&
\hspace{-0.4mm}\Delta_1:
\begin{pmatrix}
\times&\makebox[\widthof{$\times$}][c]{0}&\makebox[\widthof{$\times$}][c]{0}
\\
\makebox[\widthof{$\times$}][c]{0}&\times&\makebox[\widthof{$\times$}][c]{0}
\\
\makebox[\widthof{$\times$}][c]{0}&\makebox[\widthof{$\times$}][c]{0}&\times
\end{pmatrix},
\;
\Delta_2: 
\begin{pmatrix}
\makebox[\widthof{$\times$}][c]{0}&\makebox[\widthof{$\times$}][c]{0}&\makebox[\widthof{$\times$}][c]{0}
\\
\makebox[\widthof{$\times$}][c]{0}&\makebox[\widthof{$\times$}][c]{0}&\times
\\
\times&\makebox[\widthof{$\times$}][c]{0}&\makebox[\widthof{$\times$}][c]{0} 
\end{pmatrix},
\end{split}
\end{align*}

\vspace{-2mm}

\begin{align*}
\begin{split}
(\mathrm{Q2}):&\hspace{6.7mm}
\Gamma_1:
\begin{pmatrix}
\ast&\ast&\ast
\\
\ast&\ast&\ast
\\
\times&\times&\times
\end{pmatrix},\;\;
\Gamma_2, \Delta_2:
\begin{pmatrix}
\makebox[\widthof{$\times$}][c]{0}&\makebox[\widthof{$\times$}][c]{0}&\makebox[\widthof{$\times$}][c]{0}
\\
\makebox[\widthof{$\times$}][c]{0}&\makebox[\widthof{$\times$}][c]{0}&\makebox[\widthof{$\times$}][c]{0}
\\
\makebox[\widthof{$\times$}][c]{0}&\makebox[\widthof{$\times$}][c]{0}&\makebox[\widthof{$\times$}][c]{0} 
\end{pmatrix},
\\
&
\hspace{6.2mm}\Delta_1:
\begin{pmatrix}
\ast&\ast&\times
\\
\ast&\ast&\times
\\
\times&\times&\times
\end{pmatrix}
\end{split}
\end{align*}

\vspace{-2mm}

\begin{align*}
\begin{split}
(\mathrm{Q3}):\hspace{6.7mm}&\Gamma_1:
\begin{pmatrix}
\ast&\ast&\makebox[\widthof{$\times$}][c]{0}
\\
\ast&\ast&\makebox[\widthof{$\times$}][c]{0}
\\
\makebox[\widthof{$\times$}][c]{0}&\makebox[\widthof{$\times$}][c]{0}&\times 
\end{pmatrix},
\;\;
\Gamma_2:
\begin{pmatrix}
\makebox[\widthof{$\times$}][c]{0}&\makebox[\widthof{$\times$}][c]{0}&\makebox[\widthof{$\times$}][c]{0}
\\
\makebox[\widthof{$\times$}][c]{0}&\makebox[\widthof{$\times$}][c]{0}&\makebox[\widthof{$\times$}][c]{0}
\\
\times&\times&\makebox[\widthof{$\times$}][c]{0} 
\end{pmatrix},
\hspace{6mm}
\\
&
\hspace{-0.4mm}\Delta_1:\begin{pmatrix}
\ast&\ast&\makebox[\widthof{$\times$}][c]{0}
\\
\times&\times&\makebox[\widthof{$\times$}][c]{0}
\\
\makebox[\widthof{$\times$}][c]{0}&\makebox[\widthof{$\times$}][c]{0}&\times 
\end{pmatrix},
\;
\Delta_2:
\begin{pmatrix}
\makebox[\widthof{$\times$}][c]{0}&\makebox[\widthof{$\times$}][c]{0}&\times
\\
\makebox[\widthof{$\times$}][c]{0}&\makebox[\widthof{$\times$}][c]{0}&\times
\\
\makebox[\widthof{$\times$}][c]{0}&\makebox[\widthof{$\times$}][c]{0}&\makebox[\widthof{$\times$}][c]{0} 
\end{pmatrix},
\end{split}
\end{align*}

\vspace{-2mm}

\begin{align*}
\begin{split}
(\mathrm{Q4}):\hspace{6.7mm}&\Gamma_1:
\begin{pmatrix}
\ast&\ast&\makebox[\widthof{$\times$}][c]{0}
\\
\times&\times&\makebox[\widthof{$\times$}][c]{0}
\\
\makebox[\widthof{$\times$}][c]{0}&\makebox[\widthof{$\times$}][c]{0}&\ast 
\end{pmatrix},
\;\;
\Gamma_2:
\begin{pmatrix}
\makebox[\widthof{$\times$}][c]{0}&\makebox[\widthof{$\times$}][c]{0}&\ast
\\
\makebox[\widthof{$\times$}][c]{0}&\makebox[\widthof{$\times$}][c]{0}&\times
\\
\makebox[\widthof{$\times$}][c]{0}&\makebox[\widthof{$\times$}][c]{0}&\makebox[\widthof{$\times$}][c]{0}  
\end{pmatrix},
\hspace{6mm}
\\
&
\hspace{-0.4mm}\Delta_1:\begin{pmatrix}
\ast&\times&\makebox[\widthof{$\times$}][c]{0}
\\
\ast&\times&\makebox[\widthof{$\times$}][c]{0}
\\
\makebox[\widthof{$\times$}][c]{0}&\makebox[\widthof{$\times$}][c]{0}&\times 
\end{pmatrix},
\;
\Delta_2:
\begin{pmatrix}
\makebox[\widthof{$\times$}][c]{0}&\makebox[\widthof{$\times$}][c]{0}&\makebox[\widthof{$\times$}][c]{0}
\\
\makebox[\widthof{$\times$}][c]{0}&\makebox[\widthof{$\times$}][c]{0}&\makebox[\widthof{$\times$}][c]{0}
\\
\times&\times&\makebox[\widthof{$\times$}][c]{0}
\end{pmatrix}.
\end{split}
\end{align*}

Note that lepton texture E1 comes in one additional version, with the textures of $B$ and $C$ exchanged, and $X_S$ replaced with $-X_S$, while lepton texture E6 allows for any of the four solutions in Eq.~\eqref{ScalarPot}. 

The eight possible combinations of the lepton and quark textures are then referred to as

\begin{align}
\begin{split}
&\mathrm{Model\; A1\;to\;} \mathrm{A4}: \;\;\mathrm{E1\;with\;Q1\;to\;Q4}, 
\\
&\mathrm{Model\; A5\;to\;} \mathrm{A8}: \;\;\mathrm{E6\;with\;Q1\;to\;Q4}.
\end{split}
\end{align}

Here, each lepton texture has a maximum number of parameters that can be made real from rephasings of the charged leptons, Majorana neutrinos and Dirac neutrinos. One possible choice, which allows for the maximum number of real parameters, is shown directly in the textures above, where asterisks corresponds to complex entries and crosses to real entries. For example, lepton texture E1 has a maximum of nine parameters that can be made real simultaneously. Note that this notation applies for the entirety of Sec.~\ref{s:AnomaFree}, but not in any of the remaining sections.

Similarly, each quark texture has a maximum number of Yukawa couplings that can be made real by rephasings of the right- and left-handed quark fields. 

\subsection{Category B}

For Category B, there are two lepton textures and two quark textures, with all four combinations corresponding to valid models, model B1 to B4. Here, the lepton textures are given by

\begin{align*}
\begin{split}
(\mathrm{E2}):\;\;\;\;\;\;&\Pi_1:
\begin{pmatrix}
\ast&\ast&\makebox[\widthof{$\times$}][c]{0}
\\
\ast&\ast&\makebox[\widthof{$\times$}][c]{0}
\\
\times&\times&\makebox[\widthof{$\times$}][c]{0}
\end{pmatrix},
\;\;
\Pi_2:
\begin{pmatrix}
\makebox[\widthof{$\times$}][c]{0}&\makebox[\widthof{$\times$}][c]{0}&\ast
\\
\makebox[\widthof{$\times$}][c]{0}&\makebox[\widthof{$\times$}][c]{0}&\ast
\\
\makebox[\widthof{$\times$}][c]{0}&\makebox[\widthof{$\times$}][c]{0}&\times
\end{pmatrix},
\\
&\Sigma_1:
\begin{pmatrix}
\ast&  \makebox[\widthof{$\times$}][c]{0}
\\
\ast& \makebox[\widthof{$\times$}][c]{0}
\\
\ast & \makebox[\widthof{$\times$}][c]{0}
\end{pmatrix},
\;\;
\Sigma_2:
\begin{pmatrix}
 \makebox[\widthof{$\times$}][c]{0}& \times
\\
 \makebox[\widthof{$\times$}][c]{0}& \times
\\
 \makebox[\widthof{$\times$}][c]{0}& \times
\end{pmatrix},
\\
&A:
\begin{pmatrix}
\times& \makebox[\widthof{$\times$}][c]{0}
\\
\makebox[\widthof{$\times$}][c]{0} & \makebox[\widthof{$\times$}][c]{0} 
\end{pmatrix},
\;\;
B:
\begin{pmatrix}
\makebox[\widthof{$\times$}][c]{0} & \times
\\
\times& \makebox[\widthof{$\times$}][c]{0}
\end{pmatrix},
\;\;
C:
\begin{pmatrix}
\makebox[\widthof{$\times$}][c]{0} & \makebox[\widthof{$\times$}][c]{0}
\\
\makebox[\widthof{$\times$}][c]{0} & \makebox[\widthof{$\times$}][c]{0}
\end{pmatrix},
\end{split}
\end{align*}

\vspace{-2mm}

\begin{align*}
\begin{split}
(\mathrm{E4}):\;\;\;\;\;\;&\Pi_1:
\begin{pmatrix}
\times&\makebox[\widthof{$\times$}][c]{0}&\makebox[\widthof{$\times$}][c]{0}
\\
\makebox[\widthof{$\times$}][c]{0}&\ast&\ast
\\
\makebox[\widthof{$\times$}][c]{0} & \makebox[\widthof{$\times$}][c]{0} &  \makebox[\widthof{$\times$}][c]{0}
\end{pmatrix},
\;\;
\Pi_2:
\begin{pmatrix}
\makebox[\widthof{$\times$}][c]{0}& \makebox[\widthof{$\times$}][c]{0}&\makebox[\widthof{$\times$}][c]{0}
\\
\makebox[\widthof{$\times$}][c]{0}&\makebox[\widthof{$\times$}][c]{0}&\makebox[\widthof{$\times$}][c]{0}
\\
 \makebox[\widthof{$\times$}][c]{0}&\times&\times 
\end{pmatrix},
\\
&\Sigma_1:
\begin{pmatrix}
 \times&   \makebox[\widthof{$\times$}][c]{0}
\\
\makebox[\widthof{$\times$}][c]{0}& \makebox[\widthof{$\times$}][c]{0}
\\
\makebox[\widthof{$\times$}][c]{0} & \ast
\end{pmatrix},
\;\;
\Sigma_2:
\begin{pmatrix}
 \makebox[\widthof{$\times$}][c]{0}& \makebox[\widthof{$\times$}][c]{0}
\\
 \makebox[\widthof{$\times$}][c]{0} & \times
\\
 \times & \makebox[\widthof{$\times$}][c]{0} 
\end{pmatrix},
\\
&A:
\begin{pmatrix}
\makebox[\widthof{$\times$}][c]{0}&\makebox[\widthof{$\times$}][c]{0}
\\
\makebox[\widthof{$\times$}][c]{0}& \times
\end{pmatrix},
\;\;
B:
\begin{pmatrix}
\makebox[\widthof{$\times$}][c]{0} & \times
\\
\times & \makebox[\widthof{$\times$}][c]{0}
\end{pmatrix},
\;\;
C:
\begin{pmatrix}
\makebox[\widthof{$\times$}][c]{0} & \makebox[\widthof{$\times$}][c]{0}
\\
\makebox[\widthof{$\times$}][c]{0} & \makebox[\widthof{$\times$}][c]{0}
\end{pmatrix},
\end{split}
\end{align*}
and the two quark textures by

\begin{align*}
\begin{split}
(\mathrm{Q5}):\hspace{6mm}&\Gamma_1, \Delta_1:
\begin{pmatrix}
\ast&\ast&\makebox[\widthof{$\times$}][c]{0}
\\
\ast&\ast&\makebox[\widthof{$\times$}][c]{0}
\\
\times&\times&\makebox[\widthof{$\times$}][c]{0}
\end{pmatrix},
\;\;
\Gamma_2:
\begin{pmatrix}
\makebox[\widthof{$\times$}][c]{0}&\makebox[\widthof{$\times$}][c]{0}&\ast
\\
\makebox[\widthof{$\times$}][c]{0}&\makebox[\widthof{$\times$}][c]{0}&\ast
\\
\makebox[\widthof{$\times$}][c]{0}&\makebox[\widthof{$\times$}][c]{0}&\times
\end{pmatrix},
\\
&\hspace{5.6mm}\Delta_2:
\begin{pmatrix}
\makebox[\widthof{$\times$}][c]{0}&\makebox[\widthof{$\times$}][c]{0}&\times
\\
\makebox[\widthof{$\times$}][c]{0}&\makebox[\widthof{$\times$}][c]{0}&\times
\\
\makebox[\widthof{$\times$}][c]{0}&\makebox[\widthof{$\times$}][c]{0}&\times
\end{pmatrix},
\end{split}
\end{align*}

\vspace{-2mm}

\begin{align*}
\begin{split}
(\mathrm{Q6}):\hspace{6mm}&\Gamma_1:
\begin{pmatrix}
\ast&\ast&\ast
\\
\ast&\ast&\ast
\\
\makebox[\widthof{$\times$}][c]{0}&\makebox[\widthof{$\times$}][c]{0}&\makebox[\widthof{$\times$}][c]{0}
\end{pmatrix},
\;\;
\Gamma_2:
\begin{pmatrix}
\makebox[\widthof{$\times$}][c]{0}&\makebox[\widthof{$\times$}][c]{0}&\makebox[\widthof{$\times$}][c]{0}
\\
\makebox[\widthof{$\times$}][c]{0}&\makebox[\widthof{$\times$}][c]{0}&\makebox[\widthof{$\times$}][c]{0}
\\
\times&\times&\times
\end{pmatrix},\hspace{5.8mm}
\\
&\hspace{-0.4mm}\Delta_1:
\begin{pmatrix}
\makebox[\widthof{$\times$}][c]{0}&\makebox[\widthof{$\times$}][c]{0}&\ast
\\
\makebox[\widthof{$\times$}][c]{0}&\makebox[\widthof{$\times$}][c]{0}&\times
\\
\ast&\times&\makebox[\widthof{$\times$}][c]{0}
\end{pmatrix},
\;\;
\Delta_2:
\begin{pmatrix}
\ast&\times&\makebox[\widthof{$\times$}][c]{0}
\\
\times&\times&\makebox[\widthof{$\times$}][c]{0}
\\
\makebox[\widthof{$\times$}][c]{0}&\makebox[\widthof{$\times$}][c]{0}&\makebox[\widthof{$\times$}][c]{0} 
\end{pmatrix},
\end{split}
\end{align*}
with the four possible combinations defined as

\begin{align}
\begin{split}
&\mathrm{Model\; B1}\;(\mathrm{B2}): \;\;\mathrm{E2\;with\;Q5\;(Q6)}, 
\\
&\mathrm{Model\; B3\;}(\mathrm{B4}): \;\;\mathrm{E4\;with\;Q5\;(Q6)}.
\end{split}
\end{align}

Note also that both lepton textures come in an additional version, with $X_S$ replaced for $-X_S$ and with the textures of $B$ and $C$ exchanged. The same thing goes for the lepton texture in Category C and D.

\subsection{Category C}

For Category C, there is one lepton texture and two quark textures, with both combinations corresponding to valid models, model C1 and C2. Here, the one lepton texture is given by

\begin{align*}
\begin{split}
(\mathrm{E3}):\;\;\;\;\;\;&\Pi_1:
\begin{pmatrix}
\ast&\makebox[\widthof{$\times$}][c]{0}&\ast
\\
\makebox[\widthof{$\times$}][c]{0}&\ast&\makebox[\widthof{$\times$}][c]{0}
\\
\makebox[\widthof{$\times$}][c]{0} &\makebox[\widthof{$\times$}][c]{0}&\makebox[\widthof{$\times$}][c]{0}
\end{pmatrix},
\;\;
\Pi_2:
\begin{pmatrix}
\makebox[\widthof{$\times$}][c]{0}&\times&\makebox[\widthof{$\times$}][c]{0}
\\
\makebox[\widthof{$\times$}][c]{0} &\makebox[\widthof{$\times$}][c]{0}&\makebox[\widthof{$\times$}][c]{0}
\\
\times&\makebox[\widthof{$\times$}][c]{0}&\times
\end{pmatrix},
\\
&\Sigma_1:
\begin{pmatrix}
\ast&  \makebox[\widthof{$\times$}][c]{0}
\\
\makebox[\widthof{$\times$}][c]{0} & \makebox[\widthof{$\times$}][c]{0}
\\
\makebox[\widthof{$\times$}][c]{0} & \times
\end{pmatrix},
\;\;
\Sigma_2:
\begin{pmatrix}
 \makebox[\widthof{$\times$}][c]{0}& \times
\\
 \times&  \makebox[\widthof{$\times$}][c]{0}
\\
\makebox[\widthof{$\times$}][c]{0} & \makebox[\widthof{$\times$}][c]{0}
\end{pmatrix},
\\
&A:
\begin{pmatrix}
\times& \makebox[\widthof{$\times$}][c]{0}
\\
\makebox[\widthof{$\times$}][c]{0} & \makebox[\widthof{$\times$}][c]{0} 
\end{pmatrix},
\;\;
B:
\begin{pmatrix}
\makebox[\widthof{$\times$}][c]{0} & \times
\\
\times& \makebox[\widthof{$\times$}][c]{0}
\end{pmatrix},
\;\;
C:
\begin{pmatrix}
\makebox[\widthof{$\times$}][c]{0} & \makebox[\widthof{$\times$}][c]{0}
\\
\makebox[\widthof{$\times$}][c]{0} & \makebox[\widthof{$\times$}][c]{0}
\end{pmatrix},
\end{split}
\end{align*}
and the two quark textures by

\begin{align*}
\begin{split}
(\mathrm{Q7}):\;\;\;\;\;\;&\Gamma_1:
\begin{pmatrix}
\makebox[\widthof{$\times$}][c]{0}&\makebox[\widthof{$\times$}][c]{0}&\makebox[\widthof{$\times$}][c]{0}
\\
\makebox[\widthof{$\times$}][c]{0}&\makebox[\widthof{$\times$}][c]{0}&\ast
\\
\ast&\ast&\makebox[\widthof{$\times$}][c]{0}
\end{pmatrix},
\;\;
\Gamma_2:
\begin{pmatrix}
\times&\times&\makebox[\widthof{$\times$}][c]{0}
\\
\makebox[\widthof{$\times$}][c]{0}&\makebox[\widthof{$\times$}][c]{0}&\makebox[\widthof{$\times$}][c]{0}
\\
\makebox[\widthof{$\times$}][c]{0}&\makebox[\widthof{$\times$}][c]{0}&\times
\end{pmatrix},
\\
&\Delta_1:
\begin{pmatrix}
\makebox[\widthof{$\times$}][c]{0}&\makebox[\widthof{$\times$}][c]{0}&\times
\\
\makebox[\widthof{$\times$}][c]{0}&\makebox[\widthof{$\times$}][c]{0}&\makebox[\widthof{$\times$}][c]{0}
\\
\ast&\times&\makebox[\widthof{$\times$}][c]{0}
\end{pmatrix},
\;\;
\Delta_2:
\begin{pmatrix}
\makebox[\widthof{$\times$}][c]{0}&\makebox[\widthof{$\times$}][c]{0}&\makebox[\widthof{$\times$}][c]{0}
\\
\times&\times&\makebox[\widthof{$\times$}][c]{0}
\\
\makebox[\widthof{$\times$}][c]{0}&\makebox[\widthof{$\times$}][c]{0}&\times
\end{pmatrix},
\end{split}
\end{align*}

\vspace{-2mm}

\begin{align*}
\begin{split}
(\mathrm{Q8}):\;\;\;\;\;\;&\Gamma_1:
\begin{pmatrix}
\makebox[\widthof{$\times$}][c]{0}&\makebox[\widthof{$\times$}][c]{0}&\ast
\\
\makebox[\widthof{$\times$}][c]{0}&\makebox[\widthof{$\times$}][c]{0}&\times
\\
\ast&\ast&\makebox[\widthof{$\times$}][c]{0}
\end{pmatrix},
\;\;
\Gamma_2:
\begin{pmatrix}
\ast&\ast&\makebox[\widthof{$\times$}][c]{0}
\\
\times&\times&\makebox[\widthof{$\times$}][c]{0}
\\
\makebox[\widthof{$\times$}][c]{0}&\makebox[\widthof{$\times$}][c]{0}&\makebox[\widthof{$\times$}][c]{0}
\end{pmatrix},
\\
&\Delta_1:
\begin{pmatrix}
\ast&\ast&\times
\\
\ast&\ast&\times
\\
\makebox[\widthof{$\times$}][c]{0}&\makebox[\widthof{$\times$}][c]{0}&\makebox[\widthof{$\times$}][c]{0}
\end{pmatrix},
\;\;
\Delta_2:
\begin{pmatrix}
\makebox[\widthof{$\times$}][c]{0}&\makebox[\widthof{$\times$}][c]{0}&\makebox[\widthof{$\times$}][c]{0}
\\
\makebox[\widthof{$\times$}][c]{0}&\makebox[\widthof{$\times$}][c]{0}&\makebox[\widthof{$\times$}][c]{0}
\\
\times&\times&\times
\end{pmatrix},
\end{split}
\end{align*}
with the two possible combinations referred to as

\begin{align}
\begin{split}
&\mathrm{Model\; C1\;} (\mathrm{C2}): \;\;\mathrm{E3\;with\;Q7\;(Q8)}.
\end{split}
\end{align}

\subsection{Category D}

And finally, for Category D, there is again one lepton texture and two quark textures, with both combinations corresponding to the valid models D1 and D2. The lepton texture is given by

\begin{align*}
\begin{split}
(\mathrm{E5}):\;\;\;\;\;\;&\Pi_1:
\begin{pmatrix}
\times&\makebox[\widthof{$\times$}][c]{0}&\makebox[\widthof{$\times$}][c]{0}
\\
\makebox[\widthof{$\times$}][c]{0}&\makebox[\widthof{$\times$}][c]{0}&\makebox[\widthof{$\times$}][c]{0}
\\
\makebox[\widthof{$\times$}][c]{0} &\ast&\makebox[\widthof{$\times$}][c]{0}
\end{pmatrix},
\;\;
\Pi_2:
\begin{pmatrix}
\makebox[\widthof{$\times$}][c]{0} &\times&\makebox[\widthof{$\times$}][c]{0}
\\
\makebox[\widthof{$\times$}][c]{0} &\makebox[\widthof{$\times$}][c]{0}&\times
\\
\makebox[\widthof{$\times$}][c]{0}&\makebox[\widthof{$\times$}][c]{0}&\makebox[\widthof{$\times$}][c]{0}
\end{pmatrix},
\\
&\Sigma_1:
\begin{pmatrix}
 \times&  \makebox[\widthof{$\times$}][c]{0}
\\
\makebox[\widthof{$\times$}][c]{0} & \makebox[\widthof{$\times$}][c]{0}
\\
\makebox[\widthof{$\times$}][c]{0} &\ast
\end{pmatrix},
\;\;
\Sigma_2:
\begin{pmatrix}
\makebox[\widthof{$\times$}][c]{0} & \makebox[\widthof{$\times$}][c]{0}
\\
 \makebox[\widthof{$\times$}][c]{0}& \times
\\
 \times&  \makebox[\widthof{$\times$}][c]{0}
\end{pmatrix},
\\
&A:
\begin{pmatrix}
\makebox[\widthof{$\times$}][c]{0}& \makebox[\widthof{$\times$}][c]{0}
\\
\makebox[\widthof{$\times$}][c]{0} &  \times
\end{pmatrix},
\;\;
B:
\begin{pmatrix}
\makebox[\widthof{$\times$}][c]{0} & \times
\\
\times& \makebox[\widthof{$\times$}][c]{0}
\end{pmatrix},
\;\;
C:
\begin{pmatrix}
\makebox[\widthof{$\times$}][c]{0} & \makebox[\widthof{$\times$}][c]{0}
\\
\makebox[\widthof{$\times$}][c]{0} & \makebox[\widthof{$\times$}][c]{0}
\end{pmatrix},
\end{split}
\end{align*}
and the two quark textures by

\begin{align*}
\begin{split}
(\mathrm{Q9}):\;\;\;\;\;\;&\Gamma_1:
\begin{pmatrix}
\ast&\makebox[\widthof{$\times$}][c]{0}&\makebox[\widthof{$\times$}][c]{0}
\\
\times&\makebox[\widthof{$\times$}][c]{0}&\makebox[\widthof{$\times$}][c]{0}
\\
\makebox[\widthof{$\times$}][c]{0}&\makebox[\widthof{$\times$}][c]{0}&\times
\end{pmatrix},
\;\;
\Gamma_2:
\begin{pmatrix}
\makebox[\widthof{$\times$}][c]{0}&\ast&\makebox[\widthof{$\times$}][c]{0}
\\
\makebox[\widthof{$\times$}][c]{0}&\times&\makebox[\widthof{$\times$}][c]{0}
\\
\times&\makebox[\widthof{$\times$}][c]{0}&\makebox[\widthof{$\times$}][c]{0}
\end{pmatrix},
\\
&\Delta_1:
\begin{pmatrix}
\ast&\times&\makebox[\widthof{$\times$}][c]{0}
\\
\times&\times&\makebox[\widthof{$\times$}][c]{0}
\\
\makebox[\widthof{$\times$}][c]{0}&\makebox[\widthof{$\times$}][c]{0}&\makebox[\widthof{$\times$}][c]{0}
\end{pmatrix},
\;\;
\Delta_2:
\begin{pmatrix}
\makebox[\widthof{$\times$}][c]{0}&\makebox[\widthof{$\times$}][c]{0}&\makebox[\widthof{$\times$}][c]{0}
\\
\makebox[\widthof{$\times$}][c]{0}&\makebox[\widthof{$\times$}][c]{0}&\makebox[\widthof{$\times$}][c]{0}
\\
\makebox[\widthof{$\times$}][c]{0}&\makebox[\widthof{$\times$}][c]{0}&\times
\end{pmatrix},
\end{split}
\end{align*}

\vspace{-2mm}

\begin{align*}
\begin{split}
(\mathrm{Q10}):\;\;\;\;\;\;&\Gamma_1:
\begin{pmatrix}
\makebox[\widthof{$\times$}][c]{0}&\makebox[\widthof{$\times$}][c]{0}&\ast
\\
\makebox[\widthof{$\times$}][c]{0}&\makebox[\widthof{$\times$}][c]{0}&\times
\\
\times&\times&\makebox[\widthof{$\times$}][c]{0}
\end{pmatrix},
\;\;
\Gamma_2:
\begin{pmatrix}
\ast&\ast&\makebox[\widthof{$\times$}][c]{0}
\\
\ast&\ast&\makebox[\widthof{$\times$}][c]{0}
\\
\makebox[\widthof{$\times$}][c]{0}&\makebox[\widthof{$\times$}][c]{0}&\makebox[\widthof{$\times$}][c]{0}
\end{pmatrix},
\\
&\Delta_1:
\begin{pmatrix}
\times&\makebox[\widthof{$\times$}][c]{0}&\makebox[\widthof{$\times$}][c]{0}
\\
\times&\makebox[\widthof{$\times$}][c]{0}&\makebox[\widthof{$\times$}][c]{0}
\\
\makebox[\widthof{$\times$}][c]{0}&\makebox[\widthof{$\times$}][c]{0}&\times
\end{pmatrix},
\;\;
\Delta_2:
\begin{pmatrix}
\makebox[\widthof{$\times$}][c]{0}&\ast&\makebox[\widthof{$\times$}][c]{0}
\\
\makebox[\widthof{$\times$}][c]{0}&\times&\makebox[\widthof{$\times$}][c]{0}
\\
\times&\makebox[\widthof{$\times$}][c]{0}&\makebox[\widthof{$\times$}][c]{0}
\end{pmatrix},
\end{split}
\end{align*}
with the two possible combinations defined as

\begin{align}
\begin{split}
&\mathrm{Model\; D1\;} (\mathrm{D2}): \;\mathrm{E5\;with\;Q9\;(Q10)}. 
\end{split}
\end{align}

\section{Scanning procedure and phenomenological tests}\label{s:Scan}

The model scan is divided into four subsequent parts: the scalar potential scan, the leptonic scan, the quark scan and the full scan. This separation is done to increase its efficiency - for example, as shown in the result section, half of the models can be excluded already after the leptonic scan. Below follows a summary of each stage. 

In this section, we evaluate the performance of each parameter point from the scan based on its largest individual pull. In short, each parameter point comes with a set of corresponding theoretical predictions for each observable, where the pull is defined as the difference between the experimentally measured value and the theoretical prediction, divided by the one sigma deviation of the measurement. As such, the pull is given in units of sigma, with each observable having its separate pull. When evaluating the performance of a parameter point, we use only the (magnitude of the) pull of the observable that deviates the most from its measured value, i.e.~the \emph{largest individual pull}. 

\subsection{Scalar potential scan} 
	
With the couplings of the scalar potential defined as in Ref.~\cite{Astrid:2019}, we allow for the dimensionless, quartic couplings to vary over the range $\lambda \in [10^{-10},1]$ and the dimensionful, trilinear parameter to range over $a_1 \in [-5,0]\, \text{TeV}$. The fixed input for each minimization, on the other hand, are the scalar VEVs, $\beta$ and $v_S$. These parameters are picked to lie on a $100\times100$ grid, with $\beta$-values chosen such that $\tan\beta \in [10^{-3},10^{3}]$ in log-scale and with $v_S \in [1,10^4]\,\text{TeV}$. 

For each of the 10 000 possible values for $(\beta,v_S)$, we then find values for the scalar potential parameters that optimize:

	\begin{itemize}
		\item The tree-level tadpole equations being fulfilled;
		\item The lightest massive neutral scalar having the mass of the SM Higgs boson, $m_H=125.10\pm0.14\;\mathrm{GeV}$ \cite{Tanabashi:2018oca};
		\item One of the Goldstone bosons and the Higgs field being aligned with the SM limit;
		\item The scalar contributions\footnote{The remaining contributions to the oblique parameters are incorporated at a later stage (in the full scan).} to the oblique $S$, $T$ and $U$ parameters agreeing with \cite{Tanabashi:2018oca},
		\begin{align*}
		\begin{split}
		S=0.02\pm0.10,\;T=0.07\pm0.12, \;U=0.00\pm0.09.
		\end{split}
		\end{align*}
	\end{itemize}
	
By then removing any parameter points with a largest individual pull above $2\sigma$, $97\%$ of the initial 10 000 points survive. That is, the output of this part of the scan is 9 700 points in the $(\beta,v_S)$ space.

Note that we allow for a (small) misalignment, as the alignment condition is a part of the optimization rather than being implemented as a strict constraint. The tadpole equations are also a part of the optimization procedure, as our code is constructed to handle any number of Higgs doublets and any number of complex scalar singlets, where general analytical solutions are not available. In simpler systems, where the analytical solutions are known, we carefully monitor any deviation on this part. 

	
\subsection{Leptonic scan}
	
The leptonic sector contains the dimensionless Yukawa couplings $\left|y\right|\in [10^{-10},1]$ with $\mathrm{arg}(y)\in [10^{-10},6.28]$, the dimensionful Majorana parameters in $A\in [10,10^5]\, \text{TeV}$ and the dimensionless Majorana parameters in $B,C\in [10^{-3},1]$. For each $(\beta,v_S)$ grid point that survives the minimization of the scalar sector, we then scan, for both inverted ordering (IO) and normal ordering (NO), over these ranges to fit:

	\begin{itemize}
		\item The running charged lepton masses in \cite{Xing:2007fb};
		\item The two squared mass differences for neutrinos \cite{Tanabashi:2018oca},
		\begin{align*}
		\begin{split}
		\Delta m_{21}^2&=\left(7.53\pm0.18\right)\times10^{-5}\;\mathrm{eV}^2,
		\\
		\Delta m_{32}^2&=\left(2.444\pm{0.034}\right)\times10^{-3}\;\mathrm{eV}^2\;(\mathrm{NO}),
		\\
		\Delta m_{32}^2&=\left(-2.55\pm0.04\right)\times10^{-3}\;\mathrm{eV}^2\;(\mathrm{IO}); 
		\end{split}
		\end{align*}
		\item The angles and CP phase of the PMNS mixing matrix \cite{Tanabashi:2018oca},
				\begin{align*}
		\begin{split}
		\sin^2\left(\theta_{23}\right)&=0.512^{+0.019}_{-0.022}\;(\mathrm{NO,\;oct\;I}),
		\\
		\sin^2\left(\theta_{23}\right)&=0.542^{+0.019}_{-0.022}\;(\mathrm{NO,\;oct\;II}),
		\\
		\sin^2\left(\theta_{23}\right)&=0.536^{+0.023}_{-0.028}\;(\mathrm{IO}),
		\\
		\sin^2\left(\theta_{12}\right)&=0.307\pm0.013,
		\\
		\sin^2\left(\theta_{13}\right)&=\left(2.18\pm0.07\right)\times10^{-2},
		\\
		\delta_{\mathrm{CP}}&=1.37^{+0.18}_{-0.16}\;\pi\;\mathrm{rad}
		\end{split}
		\end{align*}
	\end{itemize}  

Here, any points with an individual pull above $4\sigma$ are disregarded, which further reduces the $(\beta,v_S)$ parameter space. The percentage of $(\beta,v_S)$ values that survive this cut, in addition to the number of $\beta$ values this corresponds to, are shown in Tab.~\ref{tab:RemainingPoints}. For example, for model A1 with IO, 63$\%$ of the $(\beta,v_S)$ values survive, i.e.~6 300, which in this specific case corresponded to 93$\%$ of the $\beta$ values remaining, i.e.~93 values. 
	
In the analysis carried out in Sec.~\ref{s:Pheno}, we see that IO outperforms NO for all models.\footnote{Note that a similar behavior can be found in Ref.~\cite{Barreiros:2018ndn}, where the authors consider the Dirac- and Majorana textures in E3, E4 and E5. While Ref.~\cite{Barreiros:2018ndn} allowed for no flavor-changing interactions for the charged leptons, these contributions are anyway typically suppressed in our analysis. As such it is reasonable to expect a somewhat similar behavior.} {Thus, while NO is by no means ruled out, the remainder of the scan of focused solely on the IO case. It is however worth mentioning that NO is favoured over IO experimentally \cite{Tanabashi:2018oca}.}


\begin{table}[ht!]
	\begin{tabular}{lllll}
	\hline
		\rule{0pt}{2.8ex}{\small{Model}} \;&  {\small{Leptonic scan}} \; &  {\small{Quark scan (IO)}} \;&  {\small{Full scan (IO)}}  \\
		\rule{0pt}{2.8ex}\;\;\;\;\;\;&{\small{NO/IO (\%)}} &{\small{In/Out (\%)}} &{\small{\# In/Best-fit}}  \\ 
		\hline\hline\\
		{\small{{A1}}} &\multirow{4}{*}  {\small{{E1:\;\;$81 \;/ \;63$}}}  &  $93\; / \;91$ & $78\;215\; / \;40\;850$ \\ 
		\rule{0pt}{2.8ex}{\small{{A2}}}  &                                            & $93\; / \;0$   &  $0\; / \;0$ \\
		\rule{0pt}{2.8ex}{\small{{A3}}}  &                                            & $93\; / \;87$  & $56\;947\; / \;79\;749$  \\
		\rule{0pt}{2.8ex}{\small{{A4}}}  &                                            & $93\; / \;71$  & $51\;558\; / \;36\;644$  \\[2mm]
		\hline \\
		{\small{{A5}}} &\multirow{4}{*}  {\small{{E6:\;\;$27 \;/ \;12$}}}  &  $75\; / \;75$& $18\;420\; / \;7\;165$ \\
		\rule{0pt}{2.8ex}{\small{{A6}}}  &                                            & $75\; / \;42$ &  $7\;134\; / \;12\;069$  \\
		\rule{0pt}{2.8ex}{\small{{A7}}}  &                                            & $75\; / \;74$ &  $10\;760\; / \;7\;735$ \\
		\rule{0pt}{2.8ex}{\small{{A8}}}  &                                            & $75\; / \;55$ &  $8\;514\; / \;4\;289$  \\[2mm]
		\hline \\
		{\small{{B1}}} &\multirow{2}{*}  {\small{{E2:\;\;$71 \;/ \;39$}}}  &  $95\; / \;95$& $47\;876\; / \;54\;339$ \\
		\rule{0pt}{2.8ex}{\small{{B2}}}  &                                            & $95\; / \;88$&  $49\;209\; / \;60\;520$   \\[2mm] 
		\hline \\
		{\small{{B3}}} &\multirow{2}{*}  {\small{{E4:\;\;$31 \;/ \;29$}}}  &  $100\; / \;100$& $39\;531\; / \;0\;$ \\
		\rule{0pt}{2.8ex}{\small{{B4}}}  &                                            & $100\; / \;88$&  $33\;991\; / \;0\;$   \\[2mm]
		\hline \\
		{\small{{C1}}} &\multirow{2}{*}  {\small{{E3:\;\;$38 \;/ \;27$}}}  &  $100\; / \;89$& $30\;325\; / \;17\;835$ \\
		\rule{0pt}{2.8ex}{\small{{C2}}}  &                                            & $100\; / \;71$&  $22\;928\; / \;5\;186$   \\[2mm]
		\hline \\
		{\small{{D1}}} &\multirow{2}{*} {\small{{E5:\;\;$0 \;\;/ \;16$}}}   &  $91\; / \;90$& $14\;429\; / \;0\;$ \\
		\rule{0pt}{2.8ex}{\small{{D2}}}  &                                            & $91\; / \;91$ & $18\;338\; / \;0\;$   \\[2mm]
		\hline
	\end{tabular}
	\caption{Percentage of surviving $(\tan\beta,v_S)$ values out of the initial 10 000 after the leptonic scan (2nd column), percentage of surviving $\tan\beta$ values relative to the initial 100 before and after the quark scan (3rd column), and the number of input points and best-fit points after the full scan (4th column).}
	\label{tab:RemainingPoints}
\end{table} 
		
\subsection{Quark scan}
	
For the quark sector, only $\beta$ enter as an input, while the magnitude and argument of the Yukawa couplings are allowed to range over the same values as in the leptonic sector. The scan procedure is then finding the set of Yukawa parameters that optimize the fit of:

	\begin{itemize}
		\item The running quark masses in \cite{Xing:2007fb};
		\item The angles and CP phase of the CKM mixing matrix, parametrized in terms of the Wolfenstein parameters  \cite{Tanabashi:2018oca}
		\begin{align*}
		\begin{split}
		\lambda&=0.22453\pm0.00044,\;\;A=0.836\pm0.015, 
		\\
		\bar{\rho}&=0.122^{+0.018}_{-0.017},\;\;\bar{\eta}=0.355^{+0.012}_{-0.011}.
		\end{split}
		\end{align*}
	\end{itemize} 
	
The percentages of surviving points, after using an individual pull of $2\sigma$ as the cut, are again shown in Tab.~\ref{tab:RemainingPoints}. For model A1 with IO, this is 91 $\%$, i.e.~91 out of the 93 $\beta$ values survive the quark scan. As the two $\beta$ values that were killed off can at most correspond to 100 $v_S$ values each, model A1 with IO has a minimum of 6 100 $(\beta,v_S)$ values after the quark scan. 

Unlike in the previous scans, we save all parameter points below the $2\sigma$ limit. There can hence be several points with the same $\beta$ value, but with different values for the magnitude and argument of the Yukawa couplings. As a result, the number of $(\beta,v_S)$ parameter points can here exceed the initial 10 000. For example, model A1 with IO has 78 215 such points, used as input for the full scan. 
	
\subsection{Full scan}
	
In the full scan, we then combine the output parameters from all previous minimizations of the largest individual pulls and use them as fixed input values. The only free parameter left to adjust is hence the gauge coupling of $U(1)'$, which we allow for to vary in the range $g^\prime \in [5\times 10^{-4},1]$. This scan contains, on top of all sectors previously described, phenomenological constraints for:

	\begin{itemize}
		\item Electroweak observables;
		\item Meson sector observables;
		\item Collider constraints.
	\end{itemize}
	
	Here, the electroweak observables include Z-pole pseudo observables, oblique parameters, off-pole cross-sections, rare top decays, atomic parity violation, electric dipole moments and muon magnetic moments, while the meson observables involve mass splittings, kaon sector CP asymmetry, B-sector CP-violating observables, leptonic decay and radiative decay. For the collider constraints, we consider only the ones coming from direct searches of the $Z'$ boson{, as all NP scalars tend to be heavier than $Z^\prime$}. For more details, see Ref.~\cite{Astrid:2019}.
	
Besides the observables considered in Ref.~\cite{Astrid:2019}, we include two additional lepton flavor violating (LFV) observables, namely two kinds of charged lepton decay - $\ell\rightarrow \ell^\prime\gamma$ and $\ell\rightarrow 3\ell$. To evaluate the new physics (NP) contribution to $\ell\rightarrow \ell^\prime\gamma$, we begin with defining the effective Hamiltonian 

\begin{align}
\label{effHH}
	\begin{split}
\mathcal{H}_{\mathrm{eff}}\equiv{C}_R \mathcal{Q}_R+{C}_L \mathcal{Q}_L,
	\end{split}
\end{align}
with the operators, for on-shell matching, defined as 

\begin{align}
	\begin{split}
\mathcal{Q}_{R(L)}&\equiv\frac{e}{16\pi^2}\bar{\ell^\prime}\sigma^{\mu\nu}{P}_{{R}({L})}\ell{F}_{\mu\nu}. 
	\end{split}
\end{align}

\begin{figure}
    \centering
    \begin{subfigure}[b]{0.27\textwidth}
     \includegraphics[width=\textwidth]{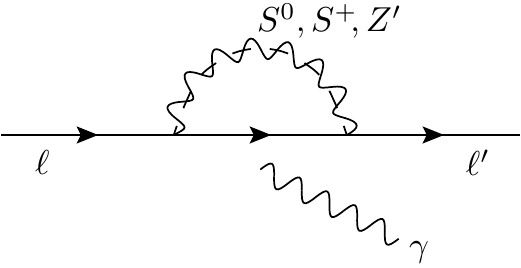}
    \end{subfigure}
    \caption{Leading order NP contribution to $\ell\rightarrow \ell^\prime\gamma$. Note that the both dashed and wiggled line is used to indicate that the propagator can be either a $Z^\prime$ boson or an NP scalar.}
    \label{fig1111}
\end{figure}

\begin{figure}
    \centering
    \begin{subfigure}[b]{0.27\textwidth}
     \includegraphics[width=\textwidth]{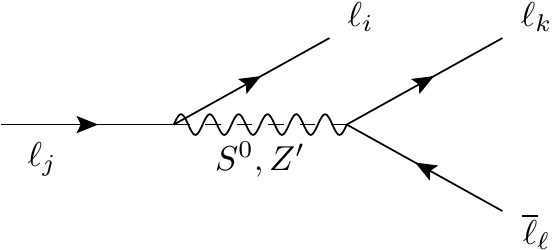}
    \end{subfigure}
    \caption{Leading order NP contribution to $\ell_j\rightarrow \ell_i\overline{\ell}_l\ell_k$.}
    \label{fig2222}
\end{figure}

From matching this (at the NP scale) to the leading order NP contributions in Fig.~\ref{fig1111}, the Wilson coefficients are given by

\begin{align}
C_{R(L)}=\frac{16\pi^2}{2ie}\frac{ {F}_2\varpm{G}_2}{m_{\ell}+m_{\ell^\prime}},
\end{align} 
where ${F}_2$ and ${G}_2$ are the so-called Pauli- and EDM form factors, calculated with (and defined as in) \textsc{Package X} \cite{Patel:2015tea}, for each parameter point. To verify the result, we compared its analytic form in the limit of massless initial- and final states with the formulae presented in Ref.~\cite{Lavoura:2003xp}. Note that the evaluation of the form factors in their exact form, i.e.~with no such limit taken, requires high precision for numerical stability, and also that the only contributions to ${F}_2$ and ${G}_2$ in Fig.~\ref{fig1111} come from diagrams where the detached photon is attached to the leptonic propagator. 


For the 3-body lepton decay, $\ell_j\rightarrow \ell_i\overline{\ell}_l\ell_k$, the leading order NP contribution is instead a tree-level diagram, as shown in Fig.~\ref{fig2222}. Here, the effective Hamiltonian (in the massless final state approximation) is given by

\begin{align}
\label{effHamil}
	\begin{split}
\mathcal{H}_{\mathrm{eff}}\equiv \;& \left[C_{V}^{XY}\right]^{ij}_{kl} \left(\overline{\ell_i}\gamma^\mu P_X\ell_{j}\right)\left(\overline{\ell_k}\gamma_\mu P_Y\ell_{l}\right)
\\
+ \;&\left[C_{S}^{XY}\right]^{ij}_{kl} \left(\overline{\ell_i} P_X\ell_{j}\right)\left(\overline{\ell_k} P_Y\ell_{l}\right),
	\end{split}
\end{align}
with an implicit summation over repeated indices and with $X,Y = L,R$. 

From matching the effective scattering amplitude to the amplitude shown in Fig.~\ref{fig2222}, we then obtain the Wilson coefficients

\begin{align}
\label{WilsonCoeff}
\begin{split}
 \left[C_{V}^{XY}\right]^{ij}_{kl} = \frac{ \Delta_X^{ij}\Delta_Y^{kl}}{M^2_{Z^\prime}} ,\;\; \left[C_{S}^{XY}\right]^{ij}_{kl} =- \frac{ \Pi_X^{ij}\Pi_Y^{kl}}{M^2_{{S^0}}},
\end{split}
\end{align}
with $\Delta$ and $\Pi$ defined as

\begin{align}
\label{Delta}
\begin{split}
\mathcal{L}=\;&\overline{\ell_i}\gamma^\mu \left(P_L\Delta_L^{ij}+ P_R\Delta_R^{ij}\right)\ell_{j}Z^\prime 
\\
+\;& \overline{\ell_i}\gamma^\mu \left(P_L\Pi_L^{ij}+ P_R\Pi_R^{ij}\right)\ell_{j}S^0, 
\end{split}
\end{align}
where $S^0$ is any neutral NP scalar and where all particles are defined to be in their mass basis. Note that, in Feynman-`t Hooft gauge, the Goldstone contribution is zero in the case of neglecting the masses of final state leptons.

The branching ratio is then given by

\begin{align}
\label{Branching}
\begin{split}
Br(\ell_j\rightarrow \ell_i\overline{\ell}_l\ell_k) = \frac{m_j^5}{N_s6144\pi^3\Gamma_{j}}\left(4N_s\left|C_{V}^{LL}\right|^2\right.
\\ 
\left.+\;4\left|C_{V}^{LR}\right|^2+N_s\left|C_{S}^{LL}\right|^2+\left|C_{S}^{LR}\right|^2\right) + \left\{L \leftrightarrow R\right\}, 
\end{split}
\end{align}
with $N_s=2$ in the case of having two identical particles in the final state (e.g.~$\tau^-\rightarrow\mu^-\mu^+\mu^-$), and $N_s=1$ otherwise, in agreement with Ref.~\cite{Crivellin:2014aa}. Here, the overall factor of $1/2$ is the phase space reduction in the case of having two identical particles in the final state, while the prefactors of $N_s$ comes from there being two possible contractions when $X=Y$ and $i=k$. For pioneering work on this subject, see Ref.~\cite{Michel_1950}.

Note that there are also contributions coming from the SM fields that mix with either $Z^\prime$ or $S^0$. This contribution is incorporated into the Wilson coefficients at the scale where the SM fields are integrated out, and have the same form as in the equations presented above. However, as this contribution is suppressed in the case of a small mixing angle, it is often negligible.

The number of best-fit points after the full scan, i.e.~number of points with the lowest individual pull for each minimization, is shown in Tab.~\ref{tab:RemainingPoints}. At this stage, any points with an individual pull above $6\sigma$ are disregarded. The reason for allowing for a larger deviation here than in the previous scans, is to properly display the shape of the distribution when plotted in Sec.~VI. Also, as there are large hadronic uncertainties in the meson sector, a model with e.g.~a $4\sigma$ deviation should not necessarily be disregarded, provided that the largest individual pull comes from an observable in the meson sector and not the EW sector.

Note also that the range for $g^\prime$ is divided into four equal parts, with each parameter point scanned for all ranges. This is why the amount of best-fit points can exceed the number of input points for some of the models in Tab.~\ref{tab:RemainingPoints}, but can never go higher than four times its value. As a rough measure of the overall performance of a model, we can hence compare the number best-fit points with four times the number of input points. For example, model A1 with IO has a ratio of $40850/(4\cdot 78215)\sim0.13$, to be compared with e.g.~$\sim0.35$ for A3 and $\sim0.18$ for A4. 

The four models with the largest ratio are A3, A6, B1 and B2, while no parameter points survive the full scan for models A2, B3, B4, D1 and D2. Note however that the ratio has a slight bias due to the various models not having exactly the same amount of statistics (the number of parameter points having survived the earlier steps in the scan of course varies in between models).

\section{Comparing the models}\label{s:Pheno}

In this section, distribution of the largest individual pulls are depicted in the form of box plots. In a box plot, the distribution is ordered by magnitude and then split into four equally sized parts, commonly referred to as quartiles. The box itself extends from the first to the third quartile, while the so-called whiskers extend out to the value with the largest magnitude within 1.5 times the length of the box in either direction. If there are any points small or large enough to be outside the extent of the whiskers, these are classified as outliers and shown as isolated black dots. The middle line corresponds to the median of the set.

Note that the largest individual pull is not necessarily a fair measure. For example, if we have a parameter point with only one deviating observable, it would be considered to perform equally to that of a parameter point where e.g.~all observables deviate with that same amount. However, it is still a useful first indicator to whether a model is promising or not.

\subsection{The leptonic sector}

\begin{figure}[ht!]
	\centering
		\begin{subfigure}[b]{0.235\textwidth}
        			\includegraphics[width=\textwidth]{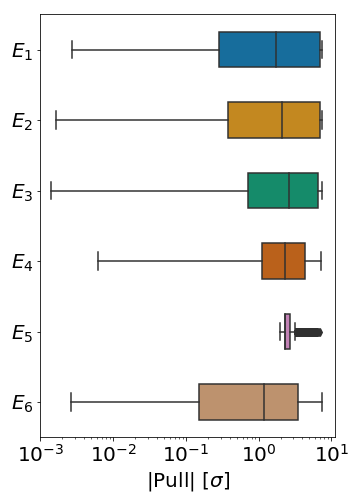}
    		\end{subfigure}
    		\begin{subfigure}[b]{0.235\textwidth}
       			\includegraphics[width=\textwidth]{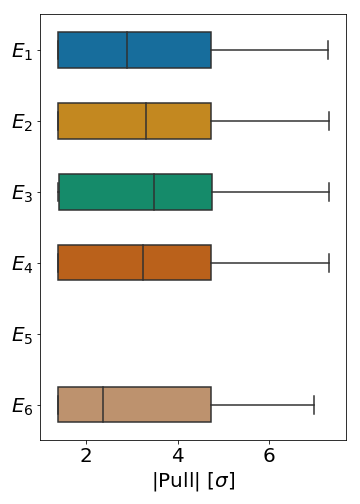}
   		\end{subfigure}
    	\caption{Distribution of the largest individual pulls for the leptonic observables, with inverted ordering (left) and normal ordering (right).} 
	\label{fig:MMLeptons}
\end{figure}

With each of the 16 models in Sec.~\ref{s:AnomaFree} coming in two varieties - one with NO and one with IO - there are a total of 32 models for us to compare. The result from the second stage of the scan, i.e.~from comparing the six lepton textures for IO and NO, is shown in Fig.~\ref{fig:MMLeptons}. Here, we see that all textures tend to perform better for IO, with the lowest individual pull ranging down to values below $10^{-2}\sigma$ for all textures but E5, rather than down to around $1.5\sigma$, as is the case for NO. In fact, for lepton texture E5, there are no valid parameter points at all for NO below the cut-off.  

\begin{figure}[ht!]
	\centering
		\begin{subfigure}[b]{0.235\textwidth}
			\includegraphics[width=\textwidth]{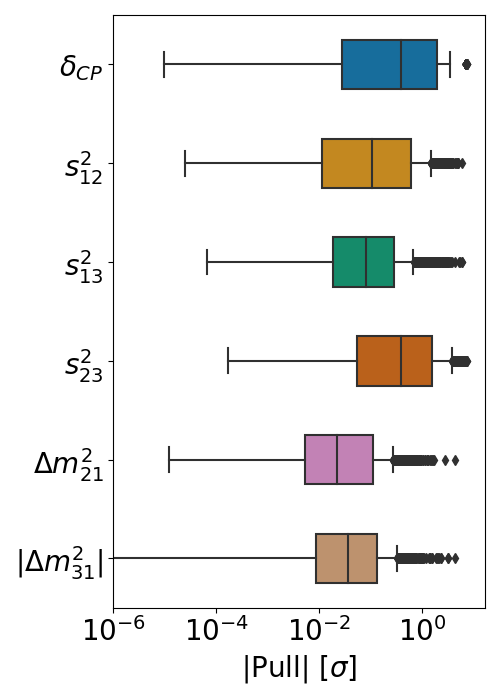}
    		\end{subfigure}
    		\begin{subfigure}[b]{0.235\textwidth}
			\includegraphics[width=\textwidth]{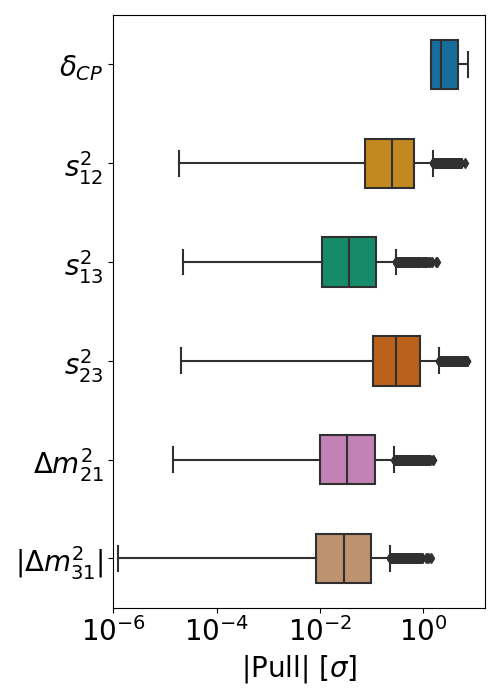}
   		\end{subfigure}
    	\caption{Distribution of pulls for model E6, with inverted ordering (left) and normal ordering (right).} 
	\label{fig:ModelE6}
\end{figure}

The source of this behavior can be identified by studying the pulls of individual observables, as shown for lepton texture E6 in Fig.~\ref{fig:ModelE6}. Here, we see that the limiting observable for NO is the CP phase, with a pull that never reaches below $2\sigma$. The same type of behavior is exercised by all of the lepton textures, which is why the quark scan and the full scan was carried out solely for lepton textures with IO, reducing the total number of models from 32 to 16.

\subsection{The full scan}

\begin{figure}[ht!]
	\centering
		\includegraphics[width=0.31\textwidth]{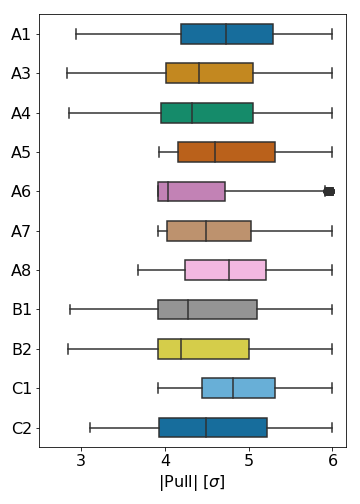}
	\caption{Distribution of the largest individual pulls, in units of sigma, for all observables mentioned in Sec.~\ref{s:Scan}. } 
	\label{fig:ALLIO}
\end{figure}

The distributions of best-fit points after the full scan are shown in Fig.~\ref{fig:ALLIO}. Note that five models are absent, namely model A2, B3, B4, D1 and D2, for which there were no parameter points with an individual pull below $7\sigma$. Overall, there are five models (A1, A3, A4, B1 and B2) with parameter points for which the largest deviation of any observable is below $3\sigma$, and neither of these points are outliers of the distribution. Among these, model A3, B1 and B2 have the largest ratio of best-fit points, as shown in Tab.~\ref{tab:RemainingPoints}. 

\section{Conclusions}\label{s:Concl}

Since the birth of 2HDMs, naturally flavor conserving implementations have been dominating the market. While they deserve the attention, their dominance has resulted in a sizable research gap. To start filling this gap, we previously classified all viable 2HDM-U(1) extensions with quarks and charged leptons in Ref.~\cite{Astrid:2019}, and, in the present work, extended the analysis to include the neutrino sector in the instance of a type-I seesaw mechanism.

The three most promising models, based on the percentage of surviving best-fit points and the distribution of the largest individual pulls, are models A3, B1 and B2, all with an inverted-hierarchical neutrino mass spectrum. 

While a scan is not a complete exploration of the parameter space, we find the results presented in this work promising. In particular, with best fit points being in the main bulk of the distribution, and not in some highly fine-tuned region of parameter space (they are not outliers), it should be relatively easy for future collaborations to find similar minima. As such, we hope that this work, and in particular model A3, B1 and B2, can prove useful for future studies.

\section*{Acknowledgements}

Work supported by the Swedish Research Council, Contract No.~2016-05996, and by the European Research Council (ERC) under the European Union Horizon 2020 research and innovation program, Grant Agreement No.~668679.

\begin{table*}[t]
		\centering
		{\small
			\begin{tabular}{p{0.004\linewidth} >{\raggedleft}p{0.15\linewidth} >{\raggedleft}p{0.15\linewidth} >{\raggedleft}p{0.14\linewidth} >{\raggedleft}p{0.13\linewidth} >{\raggedleft}p{0.15\linewidth} >{\raggedleft}p{0.14\linewidth} r}
				Model&$X_{q_L}$&$X_{u_R}$&$X_{d_R}$&$X_{\ell_L}$&$X_{e_R}$&$X_{\Phi}$&$X_{\nu_R}$\\
				\hline\hline\\
				A1&$\left[\begin{array}{c}x\\-x+2y\\y\end{array}\right]$&$\dfrac{1}{3}\left[\begin{array}{c}{4x}+{8y}\\-{2x}+{14y}\\ {x}+{11y}\end{array}\right]$
				&$\dfrac{1}{3}\left[\begin{array}{c}2x-8y\\-4x-2y\\-x-5y\end{array}\right]$&$\dfrac{1}{3}\left[\begin{array}{c}-x-8y\\-x-8y\\2x-11y\end{array}\right]$
				&$\dfrac{1}{3}\left[\begin{array}{c}-2x-16y\\-2x-16y\\x-19y\end{array}\right]$&$\dfrac{1}{3}\left[\begin{array}{c}x+8y\\4x+5y\\-3x+3y\end{array}\right]$&$\left[\begin{array}{c}0\\x-y\end{array}\right]$\\\\
				\hline\\
				A2&$x\left[\begin{array}{c}1\\1\\1\end{array}\right]$&$(x-y)\left[\begin{array}{c}1\\1\\1\end{array}\right]$
				&$(x+y)\left[\begin{array}{c}1\\1\\1\end{array}\right]$&$\left[\begin{array}{c}y\\y\\-9x-2y\end{array}\right]$
				&$\left[\begin{array}{c}2y\\2y\\-9x-y\end{array}\right]$&$\left[\begin{array}{c}-y\\-9x-4y\\9x+3y\end{array}\right]$&$\left[\begin{array}{c}0\\-9x-3y\end{array}\right]$\\\\
				\hline\\
				A3&$\left[\begin{array}{c}x\\x\\y\end{array}\right]$&$\dfrac{1}{3}\left[\begin{array}{c}{8x}+{4y}\\{8x}+{4y}\\ {5x}+{7y}\end{array}\right]$
				&$-\dfrac{1}{3}\left[\begin{array}{c}2x+4y\\2x+4y\\5x+y\end{array}\right]$&$-\dfrac{1}{3}\left[\begin{array}{c}5x+4y\\5x+4y\\8x+y\end{array}\right]$
				&$-\dfrac{1}{3}\left[\begin{array}{c}10x+8y\\10x+8y\\13x+5y\end{array}\right]$&$\dfrac{1}{3}\left[\begin{array}{c}5x+4y\\2x+7y\\3x-3y\end{array}\right]$&$\left[\begin{array}{c}0\\-x+y\end{array}\right]$\\\\
				\hline\\
				A4&$\left[\begin{array}{c}x\\x\\y\end{array}\right]$&$\dfrac{1}{3}\left[\begin{array}{c}{10x}+{2y}\\{10x}+{2y}\\ {7x}+{5y}\end{array}\right]$
				&$-\dfrac{1}{3}\left[\begin{array}{c}4x+2y\\4x+2y\\7x-y\end{array}\right]$&$-\dfrac{1}{3}\left[\begin{array}{c}7x+2y\\7x+2y\\4x+5y\end{array}\right]$
				&$-\dfrac{1}{3}\left[\begin{array}{c}14x+4y\\14x+4y\\11x+7y\end{array}\right]$&$\dfrac{1}{3}\left[\begin{array}{c}7x+2y\\10x-y\\-3x+3y\end{array}\right]$&$\left[\begin{array}{c}0\\x-y\end{array}\right]$\\\\
				\hline\\
				A5&$\left[\begin{array}{c}x\\-x+2y\\y\end{array}\right]$&$\left[\begin{array}{c}x+3y\\-x+5y\\4y\end{array}\right]$
				&$\left[\begin{array}{c}x-3y\\-x-y\\-2y\end{array}\right]$&$\left[\begin{array}{c}-3y\\-x-2y\\x-4y\end{array}\right]$
				&$\left[\begin{array}{c}-6y\\-x-5y\\x-7y\end{array}\right]$&$\left[\begin{array}{c}3y\\x+2y\\-x+y\end{array}\right]$&$\left[\begin{array}{c}0\\0\end{array}\right]$\\\\
				\hline\\
				A6&$x\left[\begin{array}{c}1\\1\\1\end{array}\right]$&$4x\left[\begin{array}{c}1\\1\\1\end{array}\right]$
				&$-2x\left[\begin{array}{c}1\\1\\1\end{array}\right]$&$\left[\begin{array}{c}-3x\\y\\-6x-y\end{array}\right]$
				&$\left[\begin{array}{c}-6x\\-3x+y\\-9x-y\end{array}\right]$&$\left[\begin{array}{c}3x\\-y\\3x+y\end{array}\right]$&$\left[\begin{array}{c}0\\0\end{array}\right]$\\\\
				\hline\\
				A7&$\left[\begin{array}{c}x\\x\\y\end{array}\right]$&$\left[\begin{array}{c}3x+y\\3x+y\\2x+2y\end{array}\right]$
				&$-\left[\begin{array}{c}x+y\\x+y\\2x\end{array}\right]$&$-\left[\begin{array}{c}2x+y\\x+2y\\3x\end{array}\right]$
				&$-\left[\begin{array}{c}4x+2y\\3x+3y\\5x+y\end{array}\right]$&$\left[\begin{array}{c}2x+y\\x+2y\\x-y\end{array}\right]$&$\left[\begin{array}{c}0\\0\end{array}\right]$\\\\
				\hline\\
				A8&$\left[\begin{array}{c}x\\x\\y\end{array}\right]$&$\left[\begin{array}{c}3x+y\\3x+y\\2x+2y\end{array}\right]$
				&$-\left[\begin{array}{c}x+y\\x+y\\2x\end{array}\right]$&$-\left[\begin{array}{c}2x+y\\3x\\x+2y\end{array}\right]$
				&$-\left[\begin{array}{c}4x+2y\\5x+y\\3x+3y\end{array}\right]$&$\left[\begin{array}{c}2x+y\\3x\\-x+y\end{array}\right]$&$\left[\begin{array}{c}0\\0\end{array}\right]$\\\\
				\hline\\
				B1&$x\left[\begin{array}{c}1\\1\\1\end{array}\right]$&$\left[\begin{array}{c}4x\\4x\\2x-y\end{array}\right]$
				&$\left[\begin{array}{c}-2x\\-2x\\y\end{array}\right]$&$-3x\left[\begin{array}{c}1\\1\\1\end{array}\right]$
				&$-\left[\begin{array}{c}6x\\6x\\4x-y\end{array}\right]$&$\left[\begin{array}{c}3x\\x-y\\2x+y\end{array}\right]$&$\left[\begin{array}{c}0\\-2x-y\end{array}\right]$\\\\
				\hline\\
				B2&$\left[\begin{array}{c}x\\x\\y\end{array}\right]$&$\left[\begin{array}{c}2x+2y\\2x+2y\\3x+y\end{array}\right]$
				&$-(x+y)\left[\begin{array}{c}1\\1\\1\end{array}\right]$&$-(2x+y)\left[\begin{array}{c}1\\1\\1\end{array}\right]$
				&$-\left[\begin{array}{c}4x+2y\\4x+2y\\3x+3y\end{array}\right]$&$\left[\begin{array}{c}2x+y\\x+2y\\x-y\end{array}\right]$&$\left[\begin{array}{c}0\\-x+y\end{array}\right]$\\\\
				\hline\\
				B3&$x\left[\begin{array}{c}1\\1\\1\end{array}\right]$&$\left[\begin{array}{c}4x\\4x\\x-y\end{array}\right]$
				&$\left[\begin{array}{c}-2x\\-2x\\x+y\end{array}\right]$&$\left[\begin{array}{c}-6x-y\\y\\-3x\end{array}\right]$
				&$-\left[\begin{array}{c}9x+y\\3x-y\\3x-y\end{array}\right]$&$\left[\begin{array}{c}3x\\-y\\3x+y\end{array}\right]$&$\left[\begin{array}{c}-3x-y\\0\end{array}\right]$\\\\
				\hline\\
				B4&$\left[\begin{array}{c}x\\x\\y\end{array}\right]$&$\left[\begin{array}{c}2x+2y\\2x+2y\\3x+y\end{array}\right]$
				&$-(x+y)\left[\begin{array}{c}1\\1\\1\end{array}\right]$&$-\left[\begin{array}{c}3x\\x+2y\\2x+y\end{array}\right]$
				&$-\left[\begin{array}{c}5x+y\\3x+3y\\3x+3y\end{array}\right]$&$\left[\begin{array}{c}2x+y\\x+2y\\x-y\end{array}\right]$&$\left[\begin{array}{c}-x+y\\0\end{array}\right]$\\\\
				\hline\hline
			\end{tabular}
		}
		\caption{Charges for model A1-A8 and B1-B4, with the additional requirement of $x\neq y$ for model A1, A3, A4, A5, A7, A8, B2 and B4, $y \neq -3x$ for model A2, A6 and B3, and $y\neq -2x$ for model B1, and with $X_{\Phi}=\{X_{\Phi_1},X_{\Phi_2},X_{S} \}$.}
		\label{tab:charges}
	\end{table*}
	
	\begin{table*}[t]
		\centering
		{\small
			\begin{tabular}{p{0.004\linewidth} >{\raggedleft}p{0.15\linewidth} >{\raggedleft}p{0.15\linewidth} >{\raggedleft}p{0.14\linewidth} >{\raggedleft}p{0.13\linewidth} >{\raggedleft}p{0.15\linewidth} >{\raggedleft}p{0.14\linewidth} r}
				Model&$q_L$&$u_R$&$d_R$&$\ell_L$&$e_R$&$\Phi$&$\nu_R$\\
				\hline\hline\\
				C1&$\left[\begin{array}{c}x\\-x+2y\\y\end{array}\right]$&$\left[\begin{array}{c}4y\\4y\\x+3y\end{array}\right]$
				&$-\left[\begin{array}{c}2y\\2y\\x+y\end{array}\right]$&$-\left[\begin{array}{c}3y\\x+2y\\-x+4y\end{array}\right]$
				&$-\left[\begin{array}{c}6y\\x+5y\\6y\end{array}\right]$&$\left[\begin{array}{c}3y\\x+2y\\-x+y\end{array}\right]$&$\left[\begin{array}{c}0\\x-y\end{array}\right]$\\\\
				\hline\\
				C2&$\left[\begin{array}{c}x\\x\\y\end{array}\right]$&$(3x+y)\left[\begin{array}{c}1\\1\\1\end{array}\right]$
				&$-\left[\begin{array}{c}2x\\2x\\x+y\end{array}\right]$&$-\left[\begin{array}{c}2x+y\\3x\\x+2y\end{array}\right]$
				&$-\left[\begin{array}{c}4x+2y\\5x+y\\4x+2y\end{array}\right]$&$\left[\begin{array}{c}2x+y\\3x\\-x+y\end{array}\right]$&$\left[\begin{array}{c}0\\x-y\end{array}\right]$\\\\
				\hline\\
				D1&$\left[\begin{array}{c}x\\x\\y\end{array}\right]$&$\left[\begin{array}{c}3x+y\\3x+y\\x+3y\end{array}\right]$
				&$-\left[\begin{array}{c}x+y\\2y\\2x\end{array}\right]$&$-\left[\begin{array}{c}3x\\x+2y\\2x+y\end{array}\right]$
				&$-\left[\begin{array}{c}5x+y\\4x+2y\\2x+4y\end{array}\right]$&$\left[\begin{array}{c}2x+y\\x+2y\\x-y\end{array}\right]$&$\left[\begin{array}{c}-x+y\\0\end{array}\right]$\\\\
				\hline\\
				D2&$\left[\begin{array}{c}x\\x\\y\end{array}\right]$&$\left[\begin{array}{c}3x+y\\4x\\2x+2y\end{array}\right]$
				&$-\left[\begin{array}{c}2x\\2x\\x+y\end{array}\right]$&$-\left[\begin{array}{c}x+2y\\3x\\2x+y\end{array}\right]$
				&$-\left[\begin{array}{c}3x+3y\\4x+2y\\6x\end{array}\right]$&$\left[\begin{array}{c}2x+y\\3x\\-x+y\end{array}\right]$&$\left[\begin{array}{c}x-y\\0\end{array}\right]$\\\\
				\hline\hline
			\end{tabular}
		}
		\caption{Charges for model C1, C2, D1 and D2, with the additional requirement of $x\neq y$ for all four models.}
		\label{tab:charges2}
	\end{table*}


\vspace{6mm}

\begin{appendix}

{
\section{Scalar potential}\label{append:scalar}

The scalar potential has the same form as in Ref.~\cite{Astrid:2019}, i.e.~it is given by $V=V_0 + \{V_1+\mathrm{h.c.}\}$, where $V_0$ is the phase insensitive part and $V_1$ the phase sensitive part, on the form of
\begin{align}
\begin{split}
V_0&=\sum_i \left(m_i^2\left|\Phi_i\right|^2+\lambda_{i}\left|\Phi_i\right|^4\right)\\
&\;\;\;\; + \lambda_{12}\left|\Phi_1\right|^2\left|\Phi_2\right|^2 + \lambda^\prime_{12}\left|\Phi_1^\dagger \Phi_2\right|^2
\\
&\;\;\;\;+m_S^2\left|S\right|^2+\lambda_S\left|S\right|^4 + \sum_i \lambda_{Si}\left|\Phi_i\right|^2\left|S\right|^2,
\end{split}
\end{align}
with $i=1,2$, and
%
\begin{align}
V_1=\begin{cases}
\hspace{0.1mm}a_{1(2)}\Phi_1^\dagger \Phi_2S^{(*)}\hspace{2.6mm}\mathrm{for}\;\;x_S=\varpm \left(x_1-x_2\right)
\\[1.2mm]
\hspace{0.1mm}a_{3(4)}\Phi_1^\dagger \Phi_2{S^{(*)}}^2\hspace{1.2mm}\mathrm{for}\;\;x_S=\varpm(x_1-x_2)/2
\end{cases}
\end{align}
respectively. Note that $x_{S}$ once again denotes the $U(1)^\prime$ charge of the scalar singlet and $x_{1,2}$ the $U(1)^\prime$ charges of the two Higgs doublets. 

The necessary and sufficient conditions for vacuum stability are then given by \cite{Kannike:2012pe} 
\begin{align}
\begin{split}
\lambda_1&\geq 0, \;\;\lambda_2\geq 0, \;\;\lambda_S\geq 0,
\\
\bar{\lambda}_{12}&\equiv \lambda_{12}+\lambda_{12}^\prime\hspace{0.5mm}\Theta(-\lambda_{12}^\prime)+2\sqrt{\lambda_1\lambda_2}\geq 0,
\\
\bar{\lambda}_{S1}&\equiv \lambda_{S1}+2\sqrt{\lambda_1\lambda_S}\geq 0,
\\
\bar{\lambda}_{S2}&\equiv \lambda_{S2}+2\sqrt{\lambda_2\lambda_S}\geq 0,
\end{split}
\end{align}
and
\begin{align}
\begin{split}
&\sqrt{\lambda_1\lambda_2\lambda_S}+\sqrt{\lambda_S}\left(\lambda_{12}+\lambda_{12}^\prime\hspace{0.5mm}\Theta(-\lambda_{12}^\prime)\right)+\lambda_{S1}\sqrt{\lambda_2}
\\
&+\lambda_{S2}\sqrt{\lambda_1}+\sqrt{\bar{\lambda}_{12}\bar{\lambda}_{S1}\bar{\lambda}_{S2}  } \geq 0,
\end{split}
\end{align}
where $\Theta(x)$ is the Heaviside step function. A sufficient (but not necessary) condition for vacuum stability is hence to use only non-negative values for the quartic couplings, which is what we use in our scan.  
}


\end{appendix}

\bibliography{bib}

\end{document}